\documentclass[10pt,twocolumn,letterpaper]{article}

\usepackage{cvpr}    
\usepackage[accsupp]{axessibility} 

\usepackage{multirow}
\usepackage{colortbl}
\usepackage{xcolor}
\usepackage{graphicx}

\usepackage{array}
\usepackage{makecell}
\usepackage{bm}

\usepackage{booktabs}
\usepackage{multirow}

\newcommand{\bll}[1]{\cellcolor{red!40}#1} 
\newcommand{\sll}[1]{\cellcolor{orange!40}#1} 
\newcommand{\tll}[1]{\cellcolor{yellow!60}#1} 
\newcommand{\wll}[1]{\cellcolor{white}#1} 

\definecolor{cvprblue}{rgb}{0.21,0.49,0.74}
\usepackage[pagebackref,breaklinks,colorlinks,allcolors=cvprblue]{hyperref}
\newcommand{\Tref}[1]{Table~\ref{#1}}
\newcommand{\eref}[1]{Eq.~\eqref{#1}}

\newcommand{\fref}[1]{Fig.~\ref{#1}}
\newcommand{\Fref}[1]{Figure~\ref{#1}}

\usepackage{xcolor}
\newcounter{todos}
\AtEndDocument{\ifnum\value{todos}>0 \PackageWarning{TODOS}{There are \arabic{todos} todos left in this paper! Fix them before submitting the paper!} \fi}

\newcommand{\real}{\mathbb{R}}

\usepackage{bm}

\newcommand{\diag}{\text{diag}}

\newcommand{\HGS}{\text{HoGS}\xspace}

\usepackage{mathtools}

\newcommand\blfootnote[1]{%
  \begingroup
  \renewcommand\thefootnote{}\footnote{#1}%
  \addtocounter{footnote}{-1}%
  \endgroup  
}
\def\paperID{10956} 
\def\confName{CVPR}
\def\confYear{2025}

\title{HoGS: Unified Near and Far Object Reconstruction \\via Homogeneous Gaussian Splatting}

\author{Xinpeng Liu$^{1\dagger}$
\qquad
Zeyi Huang$^{1\dagger}$
\qquad
Fumio Okura$^1$
\qquad
Yasuyuki Matsushita$^{1,2}$\\
$^1$The University of Osaka \qquad $^2$Microsoft Research Asia -- Tokyo\\
{\tt\small \{liu.xinpeng,huang.zeyi,okura,yasumat\}@ist.osaka-u.ac.jp}
}
\begin{document}
\twocolumn[{%
\maketitle
\begin{center}
\vspace{-4mm}
    \includegraphics[width=\linewidth]{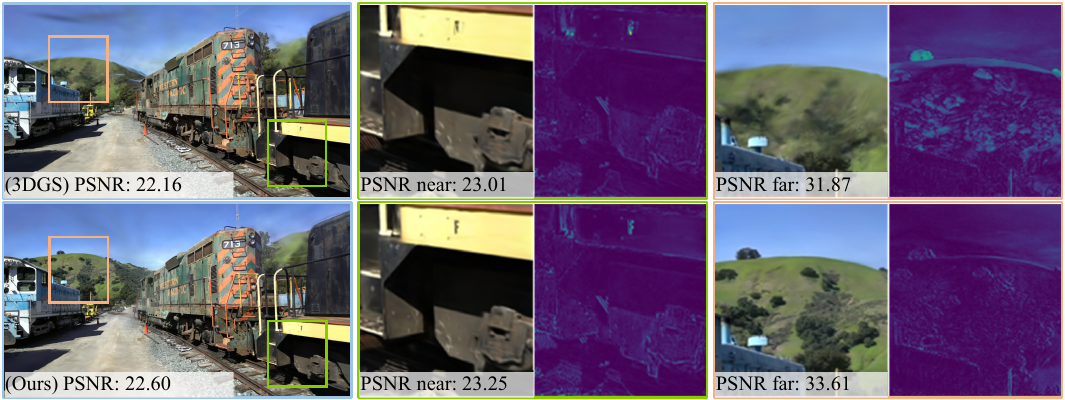}\vspace{-2mm}
    \captionof{figure}{Homogeneous Gaussian Splatting (\HGS) uses homogeneous coordinates in the 3D Gaussian Splatting (3DGS) framework to represent both positions and scales. Homogeneous coordinates defined in the projective geometry represent nearby and distant objects seamlessly, thus our \HGS enables the accurate rendering of unbounded scenes while preserving the 3DGS's fast training time and real-time performance. Compared to the traditional 3DGS, \HGS demonstrates improved robustness in capturing distant details while maintaining high fidelity in near-field objects.}
    \label{fig:intro}
\end{center}
}]

\blfootnote{$^\dagger$Authors contributed equally.}
\vspace{-1em}

\begin{abstract}
Novel view synthesis has demonstrated impressive progress recently, with 3D Gaussian splatting (3DGS) offering efficient training time and photorealistic real-time rendering. However, reliance on Cartesian coordinates limits 3DGS's performance on distant objects, which is important for reconstructing unbounded outdoor environments. We found that, despite its ultimate simplicity, using homogeneous coordinates, a concept on the projective geometry, for the 3DGS pipeline remarkably improves the rendering accuracies of distant objects.
We therefore propose Homogeneous Gaussian Splatting (\HGS) incorporating homogeneous coordinates into the 3DGS framework, providing a unified representation for enhancing near and distant objects.
\HGS effectively manages both expansive spatial positions and scales particularly in outdoor unbounded environments by adopting projective geometry principles. 
Experiments show that \HGS significantly enhances accuracy in reconstructing distant objects while maintaining high-quality rendering of nearby objects, along with fast training speed and real-time rendering capability. Our implementations are available on our project page \url{https://kh129.github.io/hogs/}.
\end{abstract}

\vspace{-5mm}
\section{Introduction}
\label{sec:intro}

Photorealistic scene reconstruction and novel view synthesis (NVS) have long been critical challenges in computer vision (CV) and computer graphics (CG). Implicit representations like neural radiance fields (NeRF)~\cite{orignerf} use neural networks to model 3D scenes as continuous functions mapping spatial coordinates and viewing directions to color and density to synthesize views using volumetric rendering. These implicit methods have achieved groundbreaking performance in NVS but are limited by computationally intensive training and slow rendering processes. 3D Gaussian Splatting (3DGS)~\cite{orig3dgs} addresses these limitations by introducing an explicit representation of scenes using 3D Gaussians, which is compatible with rasterization, accelerating training and enabling real-time rendering.

Despite these advancements, generated images by 3DGS often contain blurred parts, particularly in surrounding environments in real-world unbounded scenes, such as clouds, distant mountains, and building clusters, as shown in \fref{fig:intro}. Due to perspective projection, large objects positioned far from the camera occupy only a tiny portion of the image plane. Unbounded scenes can span arbitrarily large regions in Euclidean space, making it challenging for 3DGS to optimize Gaussian kernels to distant positions while maintaining an observable size in Euclidean space.

Inspired by the concept of traditional projective geometry and homogeneous coordinates, we introduce the homogeneous coordinates into the 3D Gaussian representation. 
We further propose a new definition of Gaussian scales that complies with the concept of projective geometry. 
Our approach leverages the advantages of homogeneous coordinates over Cartesian coordinates in representing points at infinity, providing a smooth integration of Euclidean and projective spaces while handling nearby and distant points seamlessly and effectively. Compared to 3DGS, our method significantly improves the representation of distant objects while maintaining performance a comparably high performance for nearby regions as shown in \fref{fig:intro}. 

Our contributions are twofold: first, we propose Homogeneous Gaussian Splatting (\HGS), a novel method adopting homogeneous coordinates to represent positions and scales of 3DGS for realistic and real-time rendering of both near and far objects.
Second, despite the ultimate simplicity of \HGS, our method achieves state-of-the-art NVS results compared to other implicit and explicit representations.

\section{Related Work}
\paragraph{Bounded View Synthesis.} 
We here refer to bounded view synthesis as the techniques that focus on scenes confined within a limited range~\cite{Spherical360}. All objects in the scene are within a bounded range relative to the camera, and there is no need to account for distant backgrounds.

Early NeRF methods rely on MLPs to approximate scene geometry and appearance by mapping spatial coordinates and viewing directions to density and color~\cite{orignerf}. This volumetric approach supports high-detail, multi-view-consistent reconstructions but suffers from computational intensity, as MLPs must be evaluated across numerous sampled points along each camera ray, resulting in slow rendering and limited scalability for complex scenes~\cite{neuralvox, mipnerf}. Various approaches accelerate rendering using thousands of smaller MLPs, as in KiloNeRF~\cite{kilonerf}. Techniques like InstantNGP~\cite{ingp} leverage multi-resolution hash tables for fast training and real-time rendering, albeit often with some sacrifice in image quality, particularly for high-resolution scenes. Other improvements, such as reparameterization and optimized sampling, enhance fidelity by refining ray-marching accuracy~\cite{mipnerf, pointnerf}. Despite these advancements, real-time applications remain limited by the inherent computational demands of volumetric ray-marching, a persistent bottleneck in implicit scene representations~\cite{tensorf, plenoxels}.

3DGS~\cite{orig3dgs} takes an explicit representation, \ie, modeling 3D scenes with a set of anisotropic 3D Gaussians, and achieves high-quality results through rasterization rather than ray marching. Using alpha-blending and real-time rasterization, 3DGS avoids the computational burden of NeRF's sampling strategies, producing fine details at real-time frame rates.

\vspace{-4mm}
\paragraph{Unbounded View Synthesis.} 
Several novel view synthesis studies particularly deal with unbounded views, \ie, scenes without spatial constraints, encompassing central objects and vast, often distant, backgrounds~\cite{Spherical360}.

Although implicit representations such as NeRF-based models have limitations in speed, they are effective in unbounded view synthesis by pairing specific scene types with appropriate 3D parameterizations. Several parameterization techniques for handling unbounded scenes are proposed so far. For instance, methods like NeRF++~\cite{nerfplusplus}, DONeRF~\cite{donerf}, and Mip-NeRF 360~\cite{mipnerf360, zipnerf} employ strategies called inverted sphere parameterization, radial distortion, and space contraction to model distant backgrounds, achieving realistic results despite increased computational costs. A more recent approach, Spherical Radiance Field (SRF)~\cite{Spherical360}, uses spherical structures to capture unbounded environments with concentric representations efficiently.

In contrast to the implicit representations, explicit representations like 3DGS-based models are faster but face challenges in describing distant points in unbounded Euclidean space, which often have minimal parallax and fails to reconstruct sparse 3D points using the structure-from-motion (SfM) algorithm, which is often used as the initial positions of 3D Gaussians. To address this, some methods introduce spherical initial points to model far-off elements. Skyball~\cite{gaustudio} and skybox~\cite{hierarchical} use spherically mapped points for the sky and distant areas, based on segmentation between background and foreground. Similarly, Semantics-Controlled GS (SCGS)~\cite{semanticsgs} uses scene boundaries by segmenting objects and distant areas in environments, allowing for applications such as sky replacement, while improving scene quality from diverse viewpoints. 
Scaffold-GS~\cite{scaffoldgs} employs anchor points and sparse point clouds to create a hierarchical 3D Gaussian scene representation, organizing the scene into regional anchor points and Gaussian nodes based on MLPs classifying scene types. 

Unlike the existing methods, we achieve 3DGS-based unbounded view synthesis without pre-processing defining anchor points, sky regions, or scene segmentation but just by changing the 3DGS's representation.

\section{Methods}
We propose \HGS, introducing the homogeneous projective geometry to 3DGS. 
We first summarizes the 3D Gaussian Splatting in Cartesian coordinates~\cite{orig3dgs}, and then details how we introduce homogeneous representations. 

\subsection{Cartesian 3D Gaussian Splatting}
3DGS using Cartesian coordinates is originally proposed by Kerbl~\etal~\cite{orig3dgs}, representing scenes with 3D Gaussian primitives, which are rendered with a differentiable rasterization process. Each Gaussian primitive is characterized by its 3D mean $\bm{\mu}\in\real^3$ and a covariance matrix $\bm{\Sigma}\in\real^{3\times3}$:
\begin{equation}
  {G}(\bm{p}) = \exp \left(-\frac{1}{2} (\bm{p}-\bm{\mu})^\top \bm{\Sigma}^{-1} (\bm{p}-\bm{\mu})\right),
\end{equation}
where $\bm{p}\in\real^3$ denotes a point in the 3D Cartesian coordinates. 
The covariance matrix $\bm{\Sigma}$ is decomposed into a scaling matrix $\bm{S}=\diag(s_1,s_2,s_3)$ and a rotation matrix $\bm{R}\in \mathrm{SO}(3)$ as
\begin{equation}
  \bm{\Sigma} = \bm{R}\bm{S}\bm{S}^\top\bm{R}^\top.
\end{equation}
In the 3DGS implementation, the scale and rotation are represented by a scaling vector $\bm{s}=[s_1,s_2,s_3]^\top \in\real_{\ge0}^3$ and the quaternion form $\bm{q}$, respectively. Both parameters are optimized independently in the pipeline along with the optimization of Gaussian appearances $\{o,\bm{c}\}$ containing opacity and color represented as spherical harmonics.

\vspace{-4mm}
\paragraph{Optimization and Rendering.}
To project 3D Gaussians to 2D, each Gaussian is transformed with the world-to-camera matrix $\bm{T}\in\real^{3\times4}$ and a local affine transformation $\bm{A}\in\real^{2\times3}$, from which the variance matrix of 2D Gaussian is obtained. 3DGS then applies volumetric alpha compositing to generate the rendering results.

The parameters for each 3D Gaussian, \ie, $\{\bm{\mu},\bm{s},\bm{q}\}$ for geometry and $\{o,\bm{c}\}$ for appearance, are optimized via the gradient descent approach to minimize a photometric loss combining $\mathcal{L}_{\ell 1}$ and $\mathcal{L}_{D-SSIM}$, and the optimization pipeline cooperates with an adaptive densification control to make up for missing geometric features in rendering by populating Gaussians.

\vspace{-4mm}
\paragraph{A Challenge in Cartesian 3DGS.} 
Since the original 3DGS represents the Gaussians' position $\bm{\mu}$ and scales $\bm{s}$ in the Cartesian coordinates, it theoretically requires an infinite number of iterations to represent infinitely faraway or large objects via the gradient descent optimization. Besides, the change of position parameters $\bm{\mu}$ causes the change of distance between the camera and the Gaussian primitive, affecting the size of Gaussians projected onto the image plane even if the scale parameters $\bm{s}$ does not change.
Practically, these phenomena prevent moving Gaussian primitives over large distances while preserving their shape reasonably on the projected image planes, making it difficult to reconstruct distant objects in scenes faithfully.

\subsection{Homogeneous Gaussian Splatting (\HGS)} 
We first briefly recap homogeneous coordinates, then describe how we reparametrize the position and scale of Gaussians using homogeneous coordinates.

\vspace{-4mm}
\paragraph{Homogeneous Coordinates.}
Homogeneous coordinates in 3D extend Cartesian coordinates into a four-dimensional space, which allows for a more flexible representation of geometric translation, especially for projective geometry. Let a point $\bm{p} = [x, y, z]^\top \in \mathbb{R}^3$ in 3D Cartesian coordinates is represented in projective space as $\tilde{\bm{p}} = [\tilde{x}, \tilde{y}, \tilde{z}, w]^\top \in \mathbb{P}^3$ as
\begin{equation}
\tilde{\bm{p}} = \begin{bmatrix} \tilde{x} \\ \tilde{y} \\ \tilde{z} \\ w \end{bmatrix} = w\begin{bmatrix} x \\ y \\ z \\ 1 \end{bmatrix} \propto \begin{bmatrix} \bm{p} \\ 1 \end{bmatrix}.
\label{eq:homogeneous}
\end{equation}
The fourth element $w$ is called the weight, which is often used for representing the translations and projections by a matrix-vector multiplication. 
Although not necessarily related to distance, the scaling by the weight $w$ in the homogeneous coordinates accounts for depth, ensuring further points appear smaller and respecting real-world perspective. When $w\neq0$, the homogeneous coordinates can be reduced to the Cartesian form $[\frac{\tilde{x}}{w}, \frac{\tilde{y}}{w}, \frac{\tilde{z}}{w}]^\top \in \mathbb{R}^3$. When $w=0$, the homogeneous coordinates correspond to points at infinity, essential for representing vanishing points and lines in perspective views, which are fundamental in projective geometry. This makes homogeneous coordinates an effective representation in graphics, providing a consistent framework to handle perspective transformations.

Our \HGS is inspired by the advantages of homogeneous coordinates in handling perspective projection. Leveraging homogeneous coordinates' ability to seamlessly represent points regardless of their distance, we aim to improve the reconstruction quality for distant regions while preserving the high-quality rendering of near ones.

\begin{figure}[t]
  \centering
  \includegraphics[width=\linewidth]{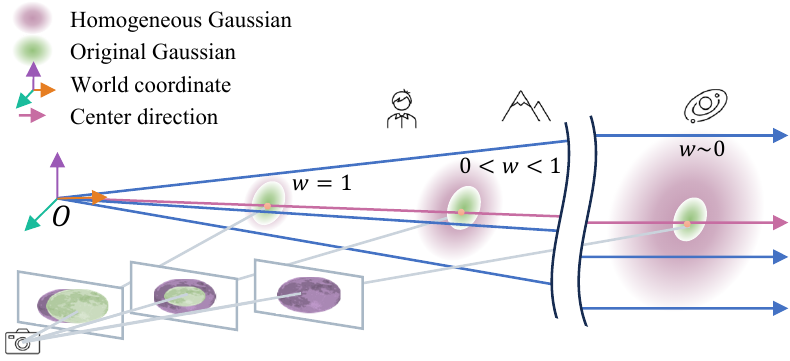}\vspace{-2mm}
   \caption{
   \textbf{Visualization of homogeneous Gaussians.}
    Our homogeneous coordinates are centered by the world origin $O$. Along the direction to the Gaussian (\ie, center direction), the distance and size of each Gaussian primitive are controlled by homogeneous position $\tilde{\bm{\mu}}\in\mathbb{P}^3$ and scaling $\tilde{\bm{s}}\in\mathbb{P}^3$. That is, even without changing the first three components in position $\{\tilde{x},\tilde{y},\tilde{z}\}$ and scaling $\{\tilde{s_1},\tilde{s_2},\tilde{s_3}\}$, the Gaussian's distance and scales change proportionally with its homogeneous component $w$. Following projective geometry principles, Gaussians further from the world origin $O$ can have smaller $w$ value to achieve proportionally larger scaling and maintain consistent projected dimensions at cameras enough close to the world origin $O$ --- a property not inherently ensured by the Cartesian 3DGS's scaling definition. Using the homogeneous scaling allows for the relative scaling of Gaussians, accurately capturing the appearance of objects at large distances. This approach simplifies the optimization of scaling for distant objects, which are often not faithfully presented in the original 3DGS's rendering.
}  \vspace{-2mm}
   \label{fig:unit_scale}
\end{figure}
\vspace{-4mm}
\paragraph{Homogeneous Gaussians.}
We here reparametrize not only positions but also scales of the Gaussian primitives in 3DGS using homogeneous coordinates. 
We consider the homogeneous coordinate system centered at the world origin $O$. Since an arbitrary 3D point in the homogeneous coordinate is given by \eref{eq:homogeneous}, the 3D mean position of Gaussians can be rewritten in its homogeneous form $\tilde{\bm{\mu}}$ as
\begin{equation}
\tilde{\bm{\mu}} = \begin{bmatrix} \bm{\mu} \\ w \end{bmatrix}\in\mathbb{P}^3 \propto \begin{bmatrix} \bm{\mu} \\ 1 \end{bmatrix},
\end{equation}
and the homogeneous mean $\tilde{\bm{\mu}}$ can be converted to its Cartesian form simply by $[\bm{\mu},1]^\top=\frac{\tilde{\bm{\mu}}}{w}$.

To unify the representation of Gaussians' scaling with their positions in homogeneous coordinates, we newly introduce the concept of \textit{homogeneous scaling}. For a Gaussian with the scaling vector $\bm{s} = [s_1, s_2, s_3]^\top \in \mathbb{R}^3$, we represent its scaling in homogeneous form as $\tilde{\mathbf{s}} = [\tilde{s_{1}}, \tilde{s_{2}}, \tilde{s_{3}}, w]^\top \in \mathbb{P}^3$, sharing the same homogeneous component $w$ as its position $\tilde{\bm{\mu}}$. Specifically, we define the homogeneous scaling as:
\begin{equation}
\tilde{\bm{s}} = \begin{bmatrix} \tilde{s_1} \\ \tilde{s_2} \\ \tilde{s_3} \\ w \end{bmatrix}\in\mathbb{P}^3 \propto \begin{bmatrix} s_1 \\ s_2 \\ s_3 \\ 1 \end{bmatrix}=\begin{bmatrix} \bm{s} \\ 1 \end{bmatrix},
\end{equation}
where $[\bm{s},1]^\top=\frac{\tilde{\bm{s}}}{w}$ as the same manner to the positions.

\Fref{fig:unit_scale} illustrates the intuition of our \HGS unifying homogeneous position $\tilde{\bm{\mu}}$ and scales $\tilde{\bm{s}}$ of Gaussian primitives. 
By incorporating the same homogeneous component $w$ into both the position and scaling vectors, we ensure that scaling operates within the same projective plane as the positions. This unified representation of position and scaling establishes a geometric framework that naturally guides the optimization of Gaussian parameters according to projective geometry principles.
Similar to how the homogeneous component $w$ in the position vector $\tilde{\bm{\mu}}$ relates to the distance from the world origin $O$ and naturally represents points at infinity, $w$ in the scale vector $\tilde{\bm{s}}$ naturally adjusts the Gaussians' size with respect to the distance. As illustrated in the figure, even with the same position $\{\tilde{x},\tilde{y},\tilde{z}\}$ and scaling $\{\tilde{s_{1}}, \tilde{s_{2}}, \tilde{s_{3}}\}$ components, Gaussians can represent distant objects with small $w\sim0$. From the cameras close enough to the origin $O$, Gaussians with different $w$ are always projected to be the same size and position on the image plane. \looseness=-1

In practice, these characteristics are useful for optimizing the scaling of objects with various depths, as homogeneous position and scaling represent the distance and size of primitives seamlessly.
The model effectively learns object scaling that generalizes well across spatial depths, as further supported by the convergence analysis in Sec.~3.3 and the experimental results in Sec.~4.

\begin{figure*}
  \centering
  \resizebox{1.00\linewidth}{!}{
  \begin{tabular}{ccc}
    \includegraphics[width=0.325\textwidth]{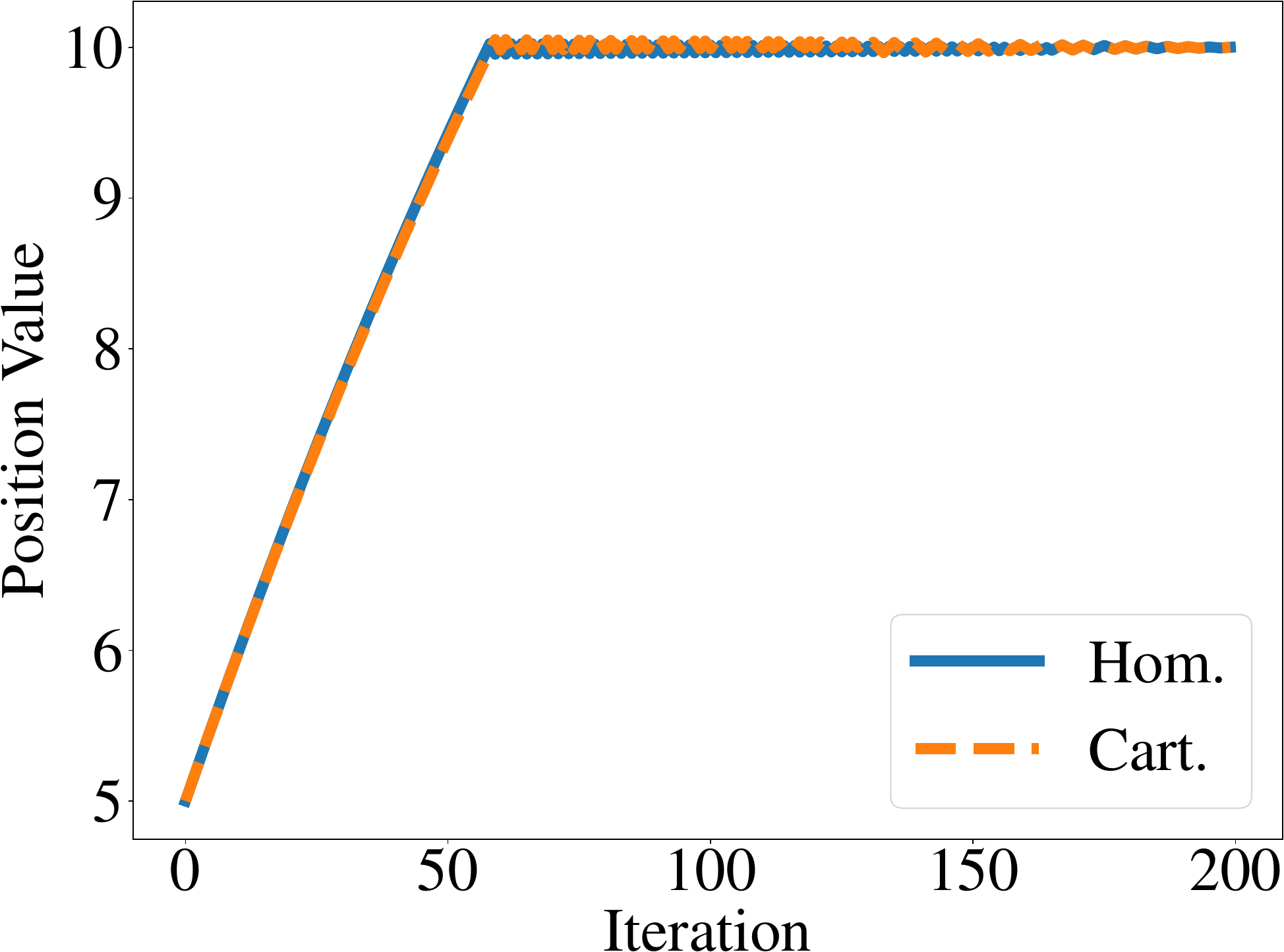} &
    \includegraphics[width=0.325\textwidth]{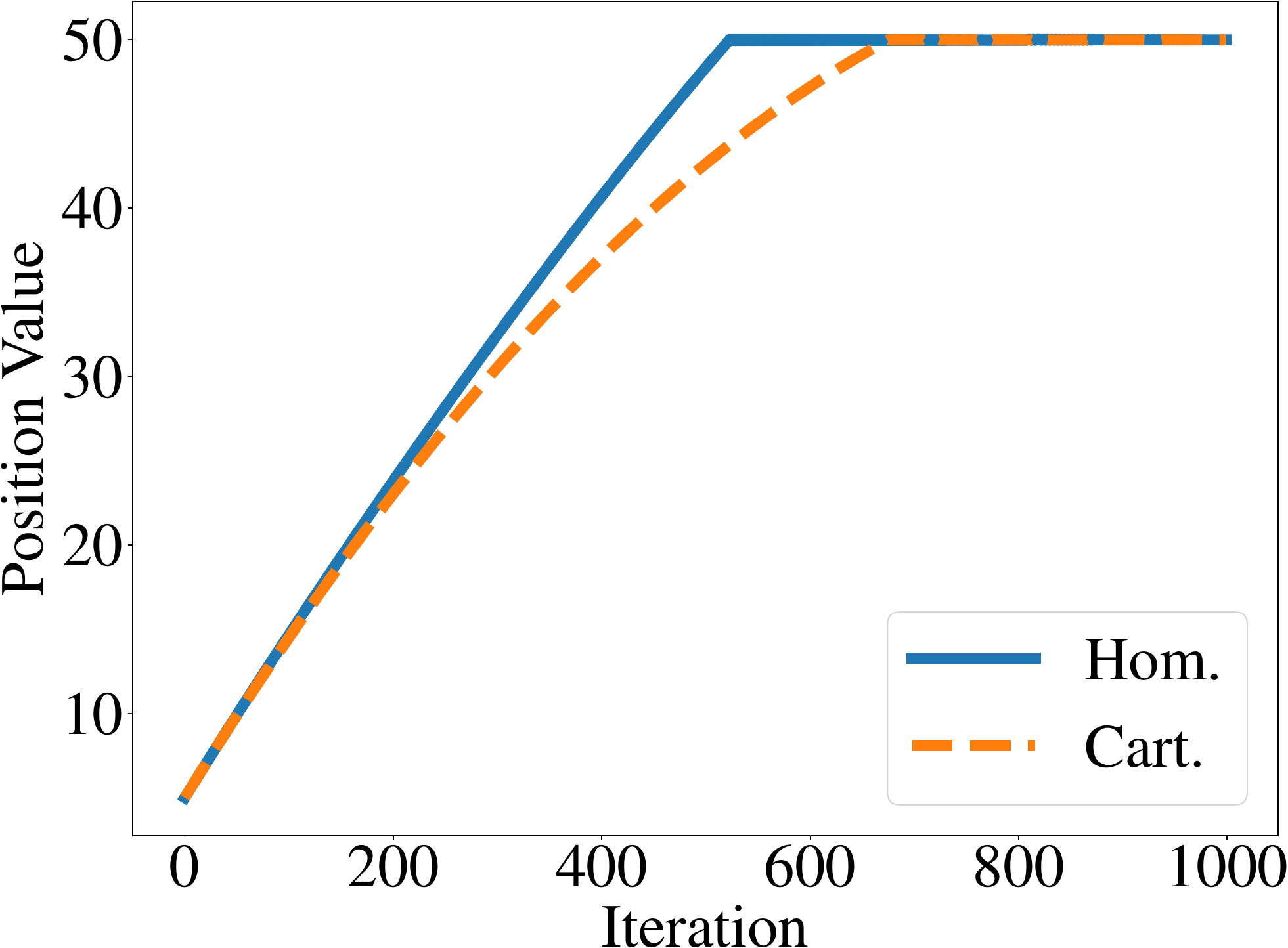} &
    \includegraphics[width=0.325\textwidth]{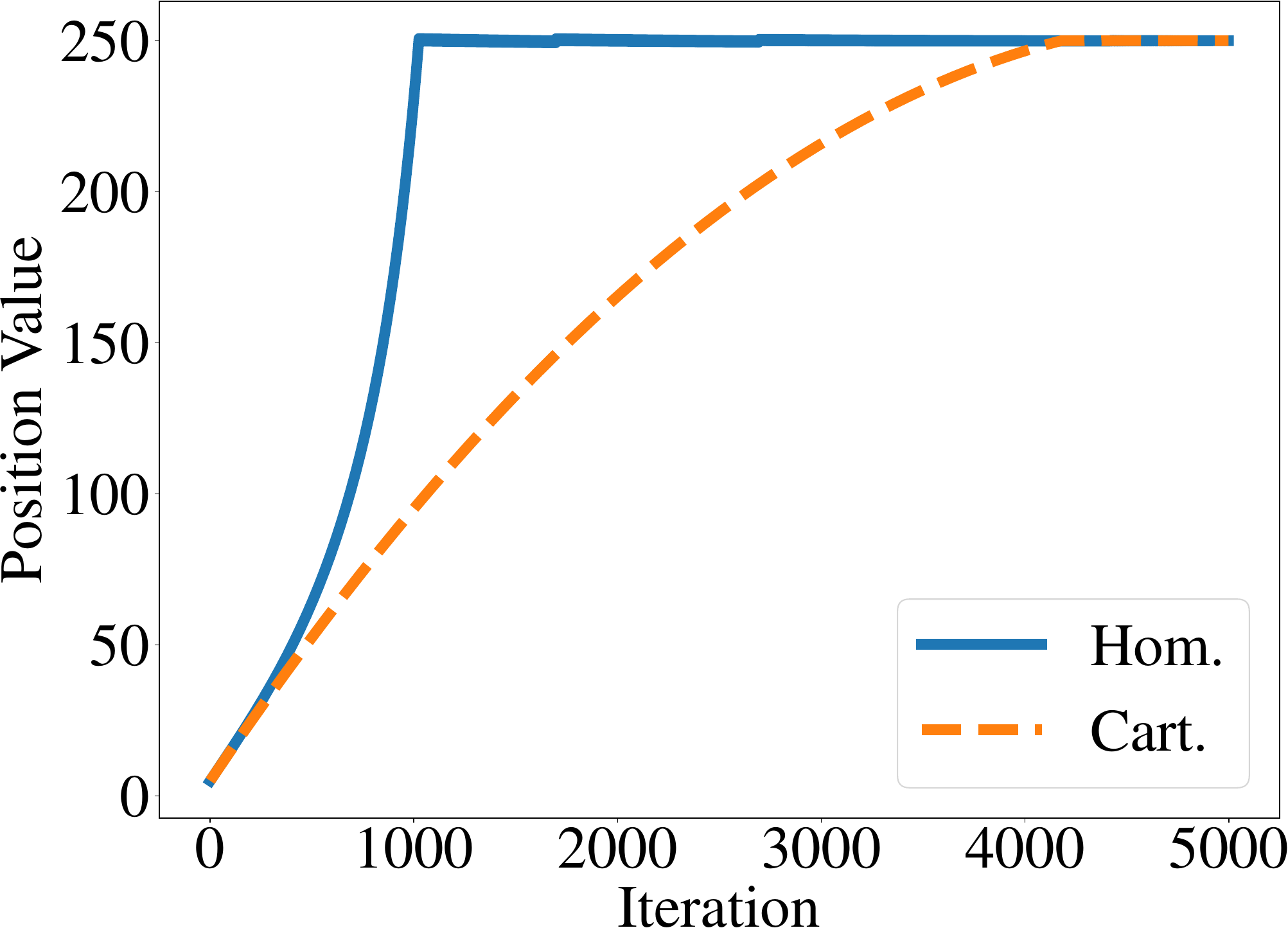} \\
    \multicolumn{1}{c}{(a) Near target ($x_t=10$)} & \multicolumn{1}{c}{(b) Intermediate target ($x_t=50$)} & \multicolumn{1}{c}{(c) Distant target ($x_t=250$)}
  \end{tabular}
}\vspace{-2mm}
\caption{
\textbf{Convergence analysis of homogeneous and Cartesian representation in 1D case.} We show a comparison of homogeneous and Cartesian representations' optimization processes based on gradient descent in the 1D simulation. For a relatively near target, both representations are close in performance, while in other cases, the homogeneous representation is capable of converging at a faster speed.}\vspace{-2mm}
\label{fig:sim1d}
\end{figure*}

\vspace{-4mm}
\paragraph{Optimization and Rendering.}
In adaptive control of Gaussians, we turn off the functionality in the original 3DGS that large Gaussians in world space need to be periodically removed throughout the optimization, so that large Gaussians with distant positions can remain in scenes. The remaining part of the optimization and rendering pipelines are the same as the original 3DGS.

\subsection{Convergence Analysis}
Homogeneous coordinates represent both near and far points equivalently, offering a more consistent treatment across spatial positions and scales. 
This approach is expected to enhance optimization performance, while we here provide an empirical analysis of the convergence speed using a 1D synthetic setup.
In detail, we optimize a specified point position relative to various target positions in a 1D space and analyze the convergence behavior of using homogeneous versus Cartesian coordinates. 
In this setup, the initial position is set to $x = 5$, which is also denoted as $\tilde{\bm{x}}=[\tilde{x}=5,w=1]^\top$ in its homogeneous coordinate.
We set multiple target positions $x_t$ (near: $x_t=10$, intermediate: $x_t=50$, and distant: $x_t=200$). 

We optimize $x$ and $\tilde{\bm{x}}$ in each representation to minimize the distance between the current and target position as
\begin{equation}
\begin{array}{ll}
  \text{Cartesian: }  &\mathcal{L}_{c} = \left|x_t - x\right|,\\
  \text{Homogeneous: }&\mathcal{L}_{h} = \left|x_t - \frac{\tilde{x}}{w}\right|,
\end{array}
\end{equation}
using the same learning rate. We observe both coordinate systems' convergence rate and final error, training each for sufficient epochs to allow convergence. \Fref{fig:sim1d} shows the results. Both representations similarly converge to the near target. However, for the intermediate and distant targets, the homogeneous representation achieved convergence in fewer iterations compared to the Cartesian representation.

\section{Experiments}
We perform both quantitative and visual comparisons using multiple datasets, as well as ablation studies.

\subsection{Experiment Setup}

\paragraph{Implementation Details.}
We implement our method with PyTorch and 3DGS's CUDA kernels for rasterization. 
In our evaluation, we used the same hyperparameters in all scenes to ensure uniformity. All experiments were run on an RTX 8000 GPU for our method. 
Our method was trained for $50,000$ iterations, and we adjusted densification-related hyperparameters to maintain the same as in the original 3DGS. 
We empirically set the learning rate of $w$ to $0.0002$.
We assign the initial value of $w$ concerning the distance $d$ of each point from the world origin $O$ by $w = \frac{1}{d} = \frac{1}{\left \lVert \mathbf{p} \right \rVert_2}$, while we find that the initialization of $w$ does not notably affect our method's behavior, as described in our ablation study in Sec.~4.3.
We use an exponential activation function for our new parameter $w$ to obtain smooth gradients. To ensure the same behavior for position parameters, we apply the standard exponential decay scheduling for mean in 3DGS to $w$. 
We adopt the same hyperparameters and design choices as 3DGS in other parts.

\vspace{-4mm}
\paragraph{Datasets.}
We evaluate the methods with $17$ real scenes from multiple datasets. To ensure that the scenes contain diverse environments, we use both bounded indoor scenes and unbounded outdoor environments.
In particular, we test our method on the full dataset of Mip-NeRF360~\cite{mipnerf360} including nine scenes containing four indoor and five outdoor scenes, two selected outdoor scenes from the Tanks\&Temples dataset~\cite{tanks}, as well as three indoor and three outdoor scenes from DL3DV benchmark~\cite{dl3dv} dataset. 

To highlight the effectiveness of our method, we also want to evaluate the synthesis accuracy for near and far objects independently. Although it is difficult to determine the threshold of near and far scenes, we use Depth Anything V2~\cite{depthanything} to generate depth maps from the input images with original resolution. We define the farthest 5\% of depth values in these maps as distant areas, and set a threshold accordingly. This evaluation is performed on unbounded scenes containing objects at varying depths, from the Tanks\&Temples and DL3DV benchmark datasets. 

\vspace{-4mm}
\paragraph{Metrics.}
To perform consistent and meaningful error metrics analysis, we follow the methodology from Mip-NeRF 360 paper~\cite{mipnerf360}, using a train/test split for datasets and taking every eighth photo for testing. Following the practice of related literature, we downsample the images to the resolution 1.0--1.6K for a fair comparison, and use the standard SSIM, PSNR, and LPIPS metrics for performance measurement.

\begin{table*}[t!]
  \small
\resizebox{1.0\linewidth}{!}{
  \tabcolsep=0.07cm
  \begin{tabular}{@{\hspace{2mm}}c@{\hspace{5mm}}c@{\hspace{3mm}}|@{\hspace{3mm}}c@{\hspace{3mm}}c@{\hspace{3mm}}c@{\hspace{3mm}}|@{\hspace{3mm}}c@{\hspace{3mm}}c@{\hspace{3mm}}c@{\hspace{3mm}}|@{\hspace{3mm}}c@{\hspace{3mm}}c@{\hspace{3mm}}c@{\hspace{2mm}}}
  \toprule
   & Dataset & \multicolumn{3}{c|@{\hspace{3mm}}}{Mip-NeRF 360 Dataset}  & \multicolumn{3}{c|@{\hspace{3mm}}}{Tanks\&Temples} & \multicolumn{3}{c@{\hspace{5mm}}}{DL3DV-10K Benchmark}\\
   & Method{\hspace{3mm}} / {\hspace{3mm}}Metric
    & SSIM$^\uparrow$   & PSNR$^\uparrow$    & LPIPS$^\downarrow$
    & SSIM$^\uparrow$   & PSNR$^\uparrow$    & LPIPS$^\downarrow$ 
    & SSIM$^\uparrow$   & PSNR$^\uparrow$    & LPIPS$^\downarrow$\\
    \midrule
\multirow{3}{*}{NeRF-based}  
    & INGP~\cite{ingp} & 0.699 & 25.59 & 0.331 & 0.745 & 21.92 & 0.305 & 0.816 & 26.38 & 0.235 \\ 
    & Mip-NeRF 360~\cite{mipnerf360} & 0.792 & 27.69 & 0.237 & 0.763 & 22.23 & 0.255 & 0.872 & 28.70 & 0.152 \\ 
   & Zip-NeRF~\cite{zipnerf}& \bll{0.829} & \bll{28.56}     & \bll{0.187} & 0.840        & 23.64      & \bll{0.151} & \bll{0.930} & \bll{30.91} & \bll{0.084} \\
    \midrule
\multirow{3}{*}{3DGS-based}
   & 3DGS (50K iter)~\cite{orig3dgs}&   0.821         & 27.65       &  0.212   & \tll{0.845} & \tll{23.83} &     0.183   &          0.900 &       28.47 &       0.143\\
   & Scaffold-GS (50K iter)~\cite{scaffoldgs} & \tll{0.825} & \sll{28.08} & \tll{0.206} & \sll{0.854} & \bll{24.31} & \tll{0.169} & \tll{0.902} &       \tll{29.11} & \tll{0.137}\\ 
   & \HGS (Ours) &    \sll{0.828} & \tll{27.92}     & \sll{0.194} & \bll{0.858} & \sll{24.27} & \sll{0.166} & \sll{0.919} & \sll{29.93} & \sll{0.114} \\
  \bottomrule
  \end{tabular}
  }\vspace{-2mm}
  \caption{
    \textbf{Quantitative comparison of novel view synthesis results.} According to the quantitative metrics, \ie, SSIM, PSNR, and LPIPS, our ultimately simple method (\HGS) achieves the best performance among 3DGS-based methods in most cases. The rendering accuracy is comparable to Zip-NeRF~\cite{zipnerf}, which shows state-of-the-art accuracy but requires time-consuming training and rendering.
  }\vspace{-2mm}
  \label{tab:comparison}
\end{table*}

\vspace{-4mm}
\paragraph{Baselines.}
Our \HGS is compared with recent NeRF-based and 3DGS-based methods to assess performance across different settings, namely INGP~\cite{ingp}, Mip-NeRF 360~\cite{mipnerf360}, Zip-NeRF~\cite{zipnerf}, Scaffold-GS~\cite{scaffoldgs}, and 3DGS~\cite{orig3dgs}. 

Especially, Zip-NeRF~\cite{zipnerf} and Scaffold-GS~\cite{scaffoldgs}, which are respectively based on NeRF and 3DGS, are considered as the representative baselines showing their state-of-the-art performance in unbounded outdoor scenes.
 
The results of INGP~\cite{ingp} on Mip-NeRF 360 and Tanks\&Temples datasets are adopted from the previous publication of 3DGS~\cite{orig3dgs}. We adopt the results of Mip-NeRF 360~\cite{mipnerf360} on their own dataset from the original paper.
The rest of the results are generated from our run of each method's official implementation with the NeRFBaselines~\cite{nerfbaselines} framework.
For fair comparisons, all models input images with the same resolution.
INGP, Mip-NeRF 360, and Zip-NeRF are tested using their default settings.
3DGS and Scaffold-GS are trained for $50,000$ iterations, the same as ours, as we find that the larger iterations also gain their performance in unbounded scenes.

\begin{table}[t!]
\centering
\resizebox{1.0\linewidth}{!}{
\begin{tabular}{c|ccc|ccc}
\toprule

\multirow{2}{*}{Method} & \multicolumn{3}{c|}{Mip-NeRF360 (\textbf{Indoor})} & \multicolumn{3}{c}{Mip-NeRF360 (\textbf{Outdoor})} \\ 
\cline{2-7} 
 & SSIM$^\uparrow$ & PSNR$^\uparrow$ & LPIPS$^\downarrow$ & SSIM$^\uparrow$ & PSNR$^\uparrow$ & LPIPS$^\downarrow$ \\ 
\midrule
Mip-NeRF 360    & \wll{0.917} & \wll{31.72} & \wll{0.180} & \wll{0.691} & \wll{24.47} & \wll{0.283} \\
Zip-NeRF    & \wll{0.929} & \wll{32.33} & \wll{0.161} & \wll{0.748} & \wll{25.54} & \wll{0.209} \\
3DGS        & \wll{0.929} & \wll{31.24} & \wll{0.169} & \wll{0.734} & \wll{24.78} & \wll{0.246} \\
Scaffold-GS & \wll{0.934} & \wll{31.79} & \wll{0.160} & \wll{0.738} & \wll{25.11} & \wll{0.243} \\
\HGS (Ours) & \wll{0.930} & \wll{31.38} & \wll{0.163} & \wll{0.747} & \wll{25.14} & \wll{0.219} \\
\midrule
\midrule

\multirow{2}{*}{Method} & \multicolumn{3}{c|}{Tanks\&Temples (\textbf{Near})} & \multicolumn{3}{c}{Tanks\&Temples (\textbf{Far})} \\ 
\cline{2-7} 
& SSIM$^\uparrow$ & PSNR$^\uparrow$ & LPIPS$^\downarrow$ & SSIM$^\uparrow$ & PSNR$^\uparrow$ & LPIPS$^\downarrow$ \\ 
\midrule
 Mip-NeRF 360    & \wll{0.800} & \wll{23.30} & \wll{0.207} & \wll{0.966} & \wll{30.47} & \wll{0.047} \\
 Zip-NeRF    & \wll{0.873} & \wll{25.07} & \wll{0.117} & \wll{0.969} & \wll{30.41} & \wll{0.034} \\
 3DGS        & \wll{0.881} & \wll{25.64} & \wll{0.140} & \wll{0.966} & \wll{29.40} & \wll{0.042} \\
 Scaffold-GS & \wll{0.885} & \wll{25.87} & \wll{0.132} & \wll{0.971} & \wll{30.54} & \wll{0.037} \\
 \HGS (Ours) & \wll{0.883} & \wll{25.69} & \wll{0.135} & \wll{0.976} & \wll{30.77} & \wll{0.031} \\ 
\midrule
\midrule

\multirow{2}{*}{Method} & \multicolumn{3}{c|}{DL3DV-10K (\textbf{Near})} & \multicolumn{3}{c}{DL3DV-10K (\textbf{Far})} \\ 
\cline{2-7} 
& SSIM$^\uparrow$ & PSNR$^\uparrow$ & LPIPS$^\downarrow$ & SSIM$^\uparrow$ & PSNR$^\uparrow$ & LPIPS$^\downarrow$ \\ 
\midrule
 Mip-NeRF 360    & \wll{0.883} & \wll{29.25} & \wll{0.133} & \wll{0.991} & \wll{39.96} & \wll{0.016} \\
 Zip-NeRF    & \wll{0.939} & \wll{31.56} & \wll{0.071} & \wll{0.993} & \wll{41.45} & \wll{0.011} \\
 3DGS        & \wll{0.918} & \wll{29.73} & \wll{0.115} & \wll{0.985} & \wll{36.46} & \wll{0.025} \\
 Scaffold-GS & \wll{0.912} & \wll{29.80} & \wll{0.118} & \wll{0.991} & \wll{39.01} & \wll{0.016} \\
 \HGS (Ours) & \wll{0.927} & \wll{30.53} & \wll{0.098} & \wll{0.993} & \wll{40.55} & \wll{0.013} \\ 

\bottomrule
\end{tabular}
}\vspace{-2mm}
\caption{\textbf{Comparisons for indoor/outdoor and near/far scenes.} For Tanks\&Temples and DL3DV-10K datasets, our \HGS shows remarkably better accuracy than other 3DGS-based methods for far scenes and also improves the near scene rendering. For Mip-NeRF 360~\cite{mipnerf360} indoor and outdoor scenes, \HGS shows strong results for general outdoor scenes, reaching close to the time-consuming Zip-NeRF.}\vspace{-2mm}
\label{tab:near_far}
\end{table}

\begin{figure*}[tp]
  \centering
  \includegraphics[width=\linewidth]{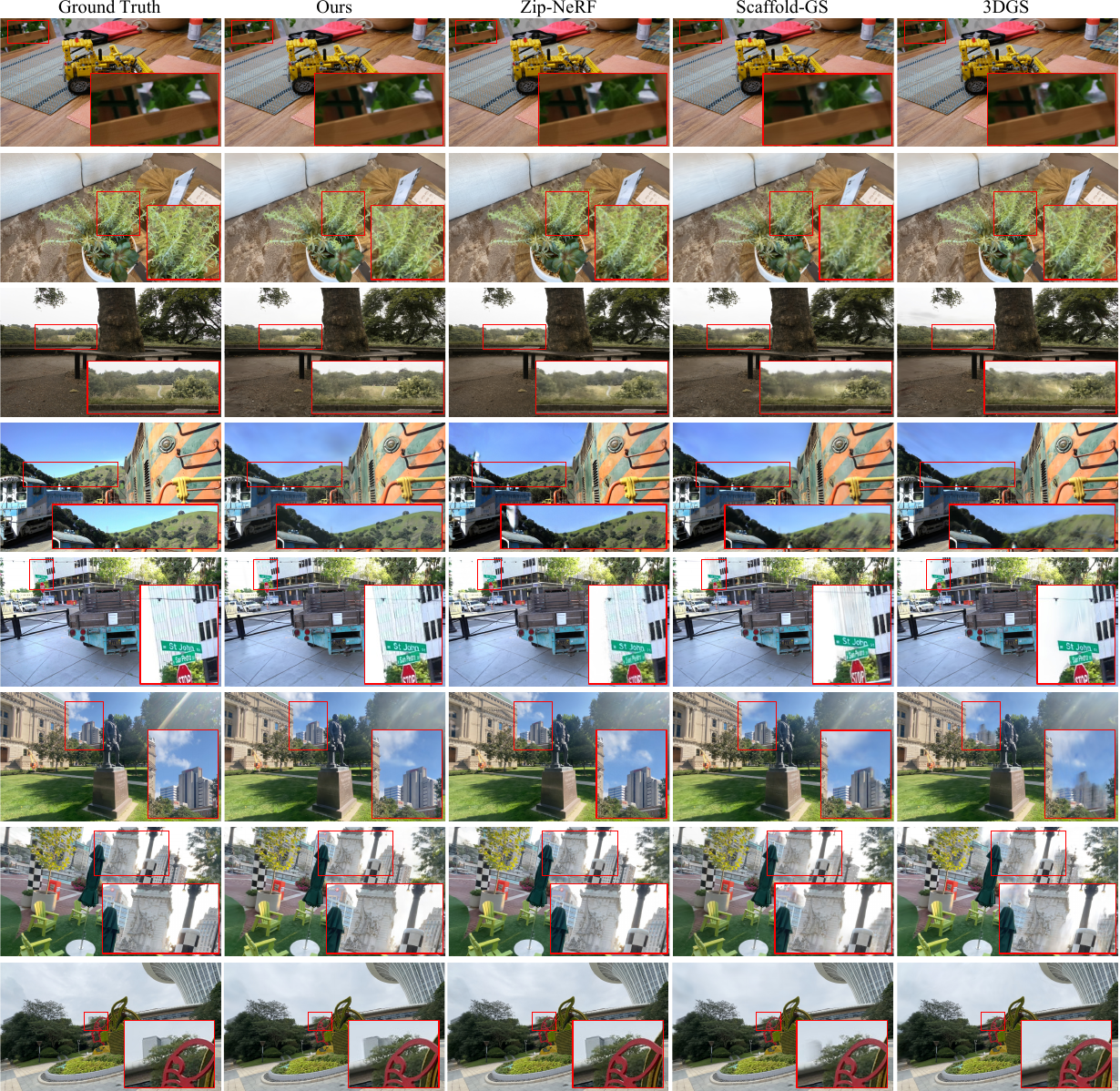}\vspace{-2mm}  
   \caption{
   \textbf{Visual comparisons on novel view synthesis.} We show comparisons of our method with previous methods on test views. The figure includes KITCHEN and TREEHILL from the Mip-NeRF360 dataset; TRAIN and TRUCK from Tanks\&Temples; and scenes 21, 26, 69, and 97 from the DL3DV benchmark. The top two rows highlight close-up details in indoor scenes, while the following rows showcase fine details in distant outdoor scenes. Key differences in quality are highlighted by insets.
   }
 \vspace{-2mm}
   \label{fig:main}
\end{figure*}


\subsection{Results and Comparisons}
We evaluated \HGS with several state-of-the-art techniques on NVS. \Tref{tab:comparison} summarizes the quantitative comparison. It demonstrates that our \HGS achieves the best rendering quality among 3DGS-based methods in most cases. 

Compared with the rendering performance of the qualitative state-of-the-art NeRF-based method (\ie, Zip-NeRF), our method often achieves comparable results and sometimes outperforms in metrics. 
Zip-NeRF requires a longer training time of an average three hours (on four A6000 GPUs), compared to 40--60 min training by ours (on a single RTX 8000 GPU). 
Since we use the same CUDA kernel to the original 3DGS, our \HGS achieves almost the same training time to 3DGS under the same number of epochs and real-time rendering. 

The qualitative analysis in ~\fref{fig:main} visually presents our advantages in rendering the quality of our method. We can see that our method reconstructs distant details, which are usually missing in 3DGS results. Furthermore, in most cases, our method preserves a comparable level of detail for objects close to the view as in 3DGS and Zip-NeRF. 
\begin{figure*}[t]
  \centering
  \includegraphics[width=\textwidth]{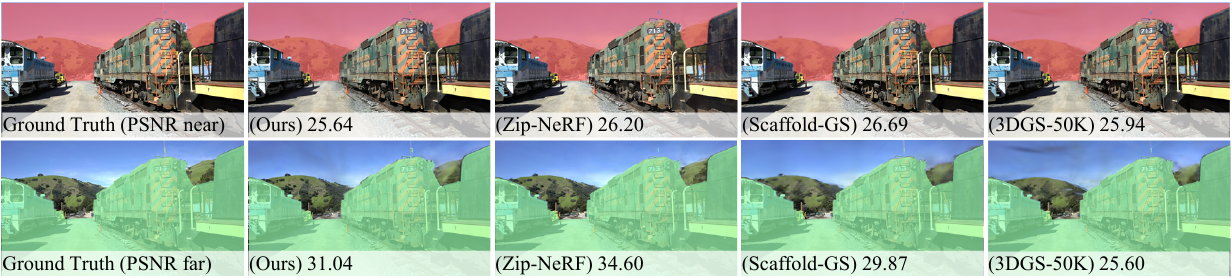}\vspace{-2.5mm}
   \caption{
   \textbf{Depth-wise visual comparison.}
    Our HoGS exhibits similar performance to 3DGS-based methods on objects close to the view, and surpasses 3DGS-based counterparts in reconstructing distant details, achieving a comparable result to Zip-NeRF.
   }
\vspace{-2mm}
   \label{fig:depth}
\end{figure*}

\vspace{-4mm}
\paragraph{Near/far and Indoor/outdoor Analyses.} We conducted detailed evaluations focusing on near and far regions (for Tanks\&Temple and DL3DV datasets) or indoor and outdoor scenes (for the Mip-NeRF360 dataset) as summarized in \Tref{tab:near_far}. 
Our \HGS demonstrates remarkably superior performance over 3DGS-based methods in representing objects across depth ranges. Notably, without explicitly exploiting scene structural information, \HGS outperforms Scaffold-GS in both near and far regions. The results for Mip-NeRF360 outdoor scenes indicate our method's strength in general outdoor scenes.
Our method achieves comparable or sometimes better results to Zip-NeRF in far regions, while maintaining a competitive performance in reconstructing near objects. 

\Fref{fig:depth} presents a visual comparison for near and far scenes, respectively. \HGS's performance is impressive in reconstructing distant background objects, preserving both geometric accuracy and detail clarity.

\subsection{Ablation Study}

\begin{figure}[t]
  \centering
  \includegraphics[width=\linewidth]{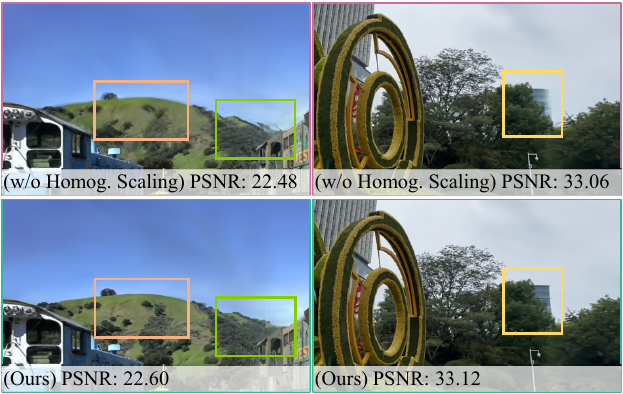}\vspace{-2mm}
\caption{
   \textbf{Ablation for homogeneous scaling.} Without homogeneous scaling, distant details become blurry as Gaussians fail to capture the precise shape of distant objects. This limitation leads to a loss of clarity in remote regions, causing indistinct contours and inaccurate representations of fine details, which diminishes the overall realism of the rendered scene.
}  
\vspace{-2mm}
   \label{fig:ablation1}
\end{figure}

\paragraph{Effect of Homogeneous Scaling.} In our method, homogeneous scaling is essential for maintaining accurate representations of near and distant objects. As shown in ~\fref{fig:ablation1}, although homogeneous representation allows our method to learn distant Gaussian positions, it is difficult for our method to learn the accurate shape of distant objects, which seem less defined in the example. 

\begin{figure}[t]
  \centering
  \includegraphics[width=\linewidth]{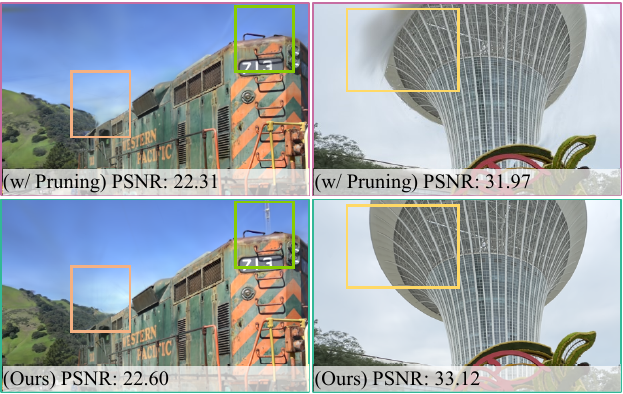}\vspace{-2mm}
   \caption{
   \textbf{Ablation for the modification on pruning strategy.} The removal of large-diameter Gaussians in world space results in the exclusion of specific distant Gaussian components. Consequently, nearby floaters approximate distant details. These floaters cannot precisely capture the structure and texture of distant regions. This substitution leads to distortions in some viewpoints, producing unrealistic renderings and insufficient detail representation in distant areas.
}  
\vspace{-2mm}
   \label{fig:ablation2}
\end{figure}

\vspace{-4mm}
\paragraph{Pruning of Large Gaussians.}
We verify the effectiveness of our modified pruning strategy for Gaussians, which allows large Gaussians in world space. In our method, distant textureless regions can be represented with large Gaussians with reasonably far positions. However, with the original pruning strategy, such large Gaussians will be removed from the scene. As a result, the method will optimize for nearby floaters to represent such regions as in ~\fref{fig:ablation2}. 

\begin{table}[tp]
\centering
\label{tab:psnr_w_init}
\resizebox{\linewidth}{!}{%
\begin{tabular}{c|c|c|c|c|c|c|c}
\hline
$w$                & $1/d$   & Random & 0.01     & 0.1    & 1.0    & 10.0   & 100.0     \\ \hline
SSIM$^\uparrow$    & 0.828  & 0.756  & 0.782    & 0.814  & 0.820  & 0.808  & 0.797  \\ 
PSNR$^\uparrow$    & 22.60  & 21.63  & 21.78    & 22.38  & 22.52  & 22.33  & 22.16  \\ 
LPIPS$^\downarrow$ & 0.191  & 0.281  & 0.241    & 0.204  & 0.204  & 0.225  & 0.240  \\ 
\hline
\end{tabular}%
}\vspace{-2mm}
\caption{Ablation experiment for different initial values of $w$. }
\label{tab:randw}
\end{table}

\vspace{-4mm}
\paragraph{Initialization of the Weight $w$.} To analyze the sensitivity of parameter $w$ to its initial value, we experimented with different initialization values for $w$ within the range of $0.01$ to $100$ and a randomly chosen value, as shown in~\Tref{tab:randw}. The results indicate negligible variation across PSNR, SSIM, and LPIPS metrics, suggesting that the initialization of $w$ has a limited impact on output quality. This robustness arises from the nature of homogeneous coordinates: a single point in Cartesian coordinates corresponds to a line through the origin in homogeneous coordinates, where any non-zero point along this line maps to the same Cartesian coordinate. Consequently, setting $w$ to $1/d$ at initialization is merely a particular solution rather than a strict requirement.

\vspace{-4mm}
\paragraph{Increasing the Number of Points.} We observe an increase in the number of points in our results compared to those produced by 3DGS. To examine the effect of the increase in the number of points, we controlled the number of points in 3DGS~\cite{orig3dgs} to match approximately that in ours (see 3DGS (Dense) in \Tref{tab:points}). Our method achieves similar performance with 3DGS (Dense) in storage, training time, and rendering FPS. Increasing the points in 3DGS does not improve its overall rendering quality and leaves its limitations in handling distant objects. 

\begin{table}[tp]
\centering
\label{tab:psnr_w_init}
\resizebox{\linewidth}{!}{%
\begin{tabular}{c|c|c|c|c|c|c}
\hline
\multirow{2}{*}{Method}&\multirow{2}{*}{\# of points}& \multirow{2}{*}{PSNR$^\uparrow$} & PSNR & Storage & Training & Rendering     \\ 
               &             &          & (\textbf{Far})$^\uparrow$ & [MB]  & [s]  & FPS     \\ \hline
3DGS           & $1,030,961$ & $26.06$ & $31.63$ & $244$  & $1,080$  & $289$   \\ \rowcolor{gray!20}
3DGS (Dense)   & $1,338,143$ & $25.98$ & $32.09$ & $319$  & $1,160$  & $212$    \\ \rowcolor{gray!20}
\HGS (Ours)    & $1,322,574$ & $27.74$ & $35.78$ & $333$  & $1,260$  & $239$   \\ 

\hline
\end{tabular}%
}
\vspace{-2.5mm}
\caption{Average performance comparisons between 3DGS and \HGS on TRAIN, and DL3DV scene 69 and 97.}
\vspace{-3.5mm}
\label{tab:points}
\end{table}

\section{Conclusions}
We have proposed Homogeneous Gaussian Splatting (\HGS), the first approach to integrating homogeneous representations into 3DGS, which provides a remarkable improvement on 3DGS via ultimately simple implementation.
Leveraging the strength of homogeneous representations, HoGS enables realistic reconstruction for both near and far objects, where original (\ie, Cartesian) 3DGS often fails to represent far objects. Despite its simplicity, HoGS's novel view rendering accuracy achieves mostly the best among 3DGS-based methods including those using complex pre-computation like scene segmentation. Rather, \HGS's accuracy nearly reaches the state-of-the-art of time-consuming NeRF-based baselines while preserving 3DGS's real-time rendering capability.

\vspace{-4mm}
\paragraph{Limitations.}
Our method partly inherits the limitation of the original 3DGS and other 3D reconstruction methods. 
It is challenging to reconstruct regions that are not well observed, leading to artifacts in the same manner as the previous methods do. 
Since our method inherits the visibility algorithm from 3DGS, it can cause abrupt changes in the depth ordering of Gaussians. 

Compared to the original 3DGS, our method preserves large Gaussians in world space and optimizes distant Gaussians to capture background details. 
This can be computationally intensive in scenes with high-frequency background details, as a result of densification. As shown in \Tref{tab:points}, our method increases the average number of points by 28.3\% compared to the original 3DGS, which lead to higher memory and storage consumption.

\vspace{-4mm}
\paragraph{Acknowledgments.}
This work was partially supported by JSPS KAKENHI Grant JP23H05491 and JST FOREST Grant JPMJFR206F.

{
    \small
    \bibliographystyle{ieeenat_fullname}
    \bibliography{main}

\begin{thebibliography}{24}
\providecommand{\natexlab}[1]{#1}
\providecommand{\url}[1]{\texttt{#1}}
\expandafter\ifx\csname urlstyle\endcsname\relax
  \providecommand{\doi}[1]{doi: #1}\else
  \providecommand{\doi}{doi: \begingroup \urlstyle{rm}\Url}\fi

\bibitem[Barron et~al.(2021)Barron, Mildenhall, Tancik, Hedman, Martin-Brualla,
  and Srinivasan]{mipnerf}
Jonathan~T. Barron, Ben Mildenhall, Matthew Tancik, Peter Hedman, Ricardo
  Martin-Brualla, and Pratul~P. Srinivasan.
\newblock {Mip-NeRF}: A multiscale representation for anti-aliasing neural
  radiance fields.
\newblock In \emph{{Proceedings of IEEE/CVF International Conference on
  Computer Vision (ICCV)}}, 2021.

\bibitem[Barron et~al.(2022)Barron, Mildenhall, Verbin, Srinivasan, and
  Hedman]{mipnerf360}
Jonathan~T. Barron, Ben Mildenhall, Dor Verbin, Pratul~P. Srinivasan, and Peter
  Hedman.
\newblock {Mip-NeRF 360}: Unbounded anti-aliased neural radiance fields.
\newblock In \emph{{Proceedings of IEEE/CVF Conference on Computer Vision and
  Pattern Recognition (CVPR)}}, 2022.

\bibitem[Barron et~al.(2023)Barron, Mildenhall, Verbin, Srinivasan, and
  Hedman]{zipnerf}
Jonathan~T. Barron, Ben Mildenhall, Dor Verbin, Pratul~P. Srinivasan, and Peter
  Hedman.
\newblock {Zip-NeRF}: Anti-aliased grid-based neural radiance fields.
\newblock In \emph{{Proceedings of IEEE/CVF International Conference on
  Computer Vision (ICCV)}}, 2023.

\bibitem[Chen et~al.(2022)Chen, Xu, Geiger, Yu, and Su]{tensorf}
Anpei Chen, Zexiang Xu, Andreas Geiger, Jingyi Yu, and Hao Su.
\newblock {TensoRF}: Tensorial radiance fields.
\newblock In \emph{{Proceedings of European Conference on Computer Vision
  (ECCV)}}, 2022.

\bibitem[Chen et~al.(2024)Chen, Wang, Lei, Dong, and Guo]{Spherical360}
Minglin Chen, Longguang Wang, Yinjie Lei, Zilong Dong, and Yulan Guo.
\newblock Learning spherical radiance field for efficient 360° unbounded novel
  view synthesis.
\newblock \emph{{IEEE Transactions on Image Processing (TIP)}}, 33:\penalty0
  3722--3734, 2024.

\bibitem[Fridovich-Keil et~al.(2022)Fridovich-Keil, Yu, Tancik, Chen, Recht,
  and Kanazawa]{plenoxels}
Sara Fridovich-Keil, Alex Yu, Matthew Tancik, Qinhong Chen, Benjamin Recht, and
  Angjoo Kanazawa.
\newblock Plenoxels: Radiance fields without neural networks.
\newblock In \emph{{Proceedings of IEEE/CVF Conference on Computer Vision and
  Pattern Recognition (CVPR)}}, 2022.

\bibitem[Kerbl et~al.(2023)Kerbl, Kopanas, Leimk{\"u}hler, and
  Drettakis]{orig3dgs}
Bernhard Kerbl, Georgios Kopanas, Thomas Leimk{\"u}hler, and George Drettakis.
\newblock {3D Gaussian} splatting for real-time radiance field rendering.
\newblock \emph{ACM Transactions on Graphics (TOG)}, 42\penalty0 (4):\penalty0
  139:1--139:14, 2023.

\bibitem[Kerbl et~al.(2024)Kerbl, Meuleman, Kopanas, Wimmer, Lanvin, and
  Drettakis]{hierarchical}
Bernhard Kerbl, Andreas Meuleman, Georgios Kopanas, Michael Wimmer, Alexandre
  Lanvin, and George Drettakis.
\newblock A hierarchical {3D} gaussian representation for real-time rendering
  of very large datasets.
\newblock \emph{ACM Transactions on Graphics (TOG)}, 43\penalty0 (4):\penalty0
  62:1--62:15, 2024.

\bibitem[Knapitsch et~al.(2017)Knapitsch, Park, Zhou, and Koltun]{tanks}
Arno Knapitsch, Jaesik Park, Qian-Yi Zhou, and Vladlen Koltun.
\newblock Tanks and temples: Benchmarking large-scale scene reconstruction.
\newblock \emph{ACM Transactions on Graphics (TOG)}, 36\penalty0 (4):\penalty0
  78:1--78:13, 2017.

\bibitem[Kulhanek and Sattler(2024)]{nerfbaselines}
Jonas Kulhanek and Torsten Sattler.
\newblock {NerfBaselines}: Consistent and reproducible evaluation of novel view
  synthesis methods.
\newblock \emph{arXiv:2406.17345}, 2024.

\bibitem[Ling et~al.(2024)Ling, Sheng, Tu, Zhao, Xin, Wan, Yu, Guo, Yu, Lu,
  et~al.]{dl3dv}
Lu Ling, Yichen Sheng, Zhi Tu, Wentian Zhao, Cheng Xin, Kun Wan, Lantao Yu,
  Qianyu Guo, Zixun Yu, Yawen Lu, et~al.
\newblock {DL3DV-10k}: A large-scale scene dataset for deep learning-based 3d
  vision.
\newblock In \emph{{Proceedings of IEEE/CVF Conference on Computer Vision and
  Pattern Recognition (CVPR)}}, 2024.

\bibitem[Liu et~al.(2020)Liu, Gu, Lin, Chua, and Theobalt]{neuralvox}
Lingjie Liu, Jiatao Gu, Kyaw~Zaw Lin, Tat-Seng Chua, and Christian Theobalt.
\newblock Neural sparse voxel fields.
\newblock In \emph{{Advances in Neural Information Processing Systems
  (NeurIPS)}}, 2020.

\bibitem[Lu et~al.(2024)Lu, Yu, Xu, Xiangli, Wang, Lin, and Dai]{scaffoldgs}
Tao Lu, Mulin Yu, Linning Xu, Yuanbo Xiangli, Limin Wang, Dahua Lin, and Bo
  Dai.
\newblock Scaffold-{GS}: Structured {3D} gaussians for view-adaptive rendering.
\newblock In \emph{{Proceedings of IEEE/CVF Conference on Computer Vision and
  Pattern Recognition (CVPR)}}, 2024.

\bibitem[Mildenhall et~al.(2020)Mildenhall, Srinivasan, Tancik, Barron,
  Ramamoorthi, and Ng]{orignerf}
Ben Mildenhall, Pratul~P. Srinivasan, Matthew Tancik, Jonathan~T. Barron, Ravi
  Ramamoorthi, and Ren Ng.
\newblock {NeRF}: Representing scenes as neural radiance fields for view
  synthesis.
\newblock In \emph{{Proceedings of European Conference on Computer Vision
  (ECCV)}}, 2020.

\bibitem[M\"uller et~al.(2022)M\"uller, Evans, Schied, and Keller]{ingp}
Thomas M\"uller, Alex Evans, Christoph Schied, and Alexander Keller.
\newblock Instant neural graphics primitives with a multiresolution hash
  encoding.
\newblock \emph{ACM Transactions on Graphics (TOG)}, 41\penalty0 (4):\penalty0
  102:1--102:15, 2022.

\bibitem[Neff et~al.(2021)Neff, Stadlbauer, Parger, Kurz, Mueller, Chaitanya,
  Kaplanyan, and Steinberger]{donerf}
Thomas Neff, Pascal Stadlbauer, Mathias Parger, Andreas Kurz, Joerg~H. Mueller,
  Chakravarty R.~Alla Chaitanya, Anton~S. Kaplanyan, and Markus Steinberger.
\newblock {DONeRF}: Towards real-time rendering of compact neural radiance
  fields using depth oracle networks.
\newblock \emph{Computer Graphics Forum}, 40\penalty0 (4):\penalty0 45--59,
  2021.

\bibitem[Reiser et~al.(2021)Reiser, Peng, Liao, and Geiger]{kilonerf}
Christian Reiser, Songyou Peng, Yiyi Liao, and Andreas Geiger.
\newblock {KiloNeRF}: Speeding up neural radiance fields with thousands of tiny
  {MLPs}.
\newblock In \emph{{Proceedings of IEEE/CVF International Conference on
  Computer Vision (ICCV)}}, 2021.

\bibitem[Schieber et~al.(2024)Schieber, Young, Langlotz, Zollmann, and
  Roth]{semanticsgs}
Hannah Schieber, Jacob Young, Tobias Langlotz, Stefanie Zollmann, and Daniel
  Roth.
\newblock Semantics-controlled gaussian splatting for outdoor scene
  reconstruction and rendering in virtual reality.
\newblock In \emph{Proceedings of IEEE Conference on Virtual Reality and 3D
  User Interfaces (VR)}, 2024.

\bibitem[Xu et~al.(2022)Xu, Xu, Philip, Bi, Shu, Sunkavalli, and
  Neumann]{pointnerf}
Qiangeng Xu, Zexiang Xu, Julien Philip, Sai Bi, Zhixin Shu, Kalyan Sunkavalli,
  and Ulrich Neumann.
\newblock {Point-NeRF}: Point-based neural radiance fields.
\newblock In \emph{{Proceedings of IEEE/CVF Conference on Computer Vision and
  Pattern Recognition (CVPR)}}, 2022.

\bibitem[Yan et~al.(2024)Yan, Low, Chen, and
  Lee]{yan2024multiscale3dgaussiansplatting}
Zhiwen Yan, Weng~Fei Low, Yu Chen, and Gim~Hee Lee.
\newblock Multi-scale {3D} gaussian splatting for anti-aliased rendering.
\newblock In \emph{{Proceedings of IEEE/CVF Conference on Computer Vision and
  Pattern Recognition (CVPR)}}, 2024.

\bibitem[Yang et~al.(2024)Yang, Kang, Huang, Zhao, Xu, Feng, and
  Zhao]{depthanything}
Lihe Yang, Bingyi Kang, Zilong Huang, Zhen Zhao, Xiaogang Xu, Jiashi Feng, and
  Hengshuang Zhao.
\newblock {Depth Anything} v2.
\newblock In \emph{{Advances in Neural Information Processing Systems
  (NeurIPS)}}, 2024.

\bibitem[Ye et~al.(2024)Ye, Nie, Chang, Chen, Zhi, and Han]{gaustudio}
Chongjie Ye, Yinyu Nie, Jiahao Chang, Yuantao Chen, Yihao Zhi, and Xiaoguang
  Han.
\newblock {GauStudio}: A modular framework for {3D} gaussian splatting and
  beyond.
\newblock \emph{arXiv:2403.19632}, 2024.

\bibitem[Yu et~al.(2024)Yu, Chen, Huang, Sattler, and
  Geiger]{Yu2024MipSplatting}
Zehao Yu, Anpei Chen, Binbin Huang, Torsten Sattler, and Andreas Geiger.
\newblock Mip-splatting: Alias-free {3D} gaussian splatting.
\newblock In \emph{{Proceedings of IEEE/CVF Conference on Computer Vision and
  Pattern Recognition (CVPR)}}, 2024.

\bibitem[Zhang et~al.(2020)Zhang, Riegler, Snavely, and Koltun]{nerfplusplus}
Kai Zhang, Gernot Riegler, Noah Snavely, and Vladlen Koltun.
\newblock {NeRF++}: Analyzing and improving neural radiance fields.
\newblock \emph{arXiv:2010.07492}, 2020.

\end{thebibliography}
}

\setcounter{page}{1}

\setcounter{figure}{0}
\setcounter{table}{0}
\setcounter{equation}{0}
\setcounter{section}{0}
\renewcommand\thefigure{S\arabic{figure}}
\renewcommand\thetable{S\arabic{table}}
\renewcommand\theequation{S\arabic{equation}}
\renewcommand\thesection{\Alph{section}}

\maketitlesupplementary

\noindent This supplementary material first provides a detailed comparison of our homogeneous representation with existing inverted spherical representation~\cite{nerfplusplus} (Sec.~\ref{supp:spherical}).
We also analyze the methods' convergence behavior (Sec.~\ref{supp:convergence}) and potential design choices in methods and experiments (Sec.~\ref{supp:design}).
An experiment in Sec.~\ref{supp:moon} assesses whether our method can represent an \emph{infinitely} far object (\ie, the Moon) in our dataset.
Finally, Sec.~\ref{supp:additional} shows additional and detailed results in each scene we tested. 

The supplementary video visualizes the results in a more intuitive manner, and we strongly encourage the reader to refer to it.

\section{Inverted Spherical vs.~Homogeneous Representations}
\label{supp:spherical}

Similar to the homogeneous representation, the inverted spherical representation in NeRF++~\cite{nerfplusplus} is also designed to represent distant objects effectively.
We compare the performance of inverted spherical and homogeneous representation in 3D scene representation. 

\paragraph{Inverted Spherical Representation~\cite{nerfplusplus}.} We define a point $\bm{p} = [x, y, z]^\top \in \mathbb{R}^3$ to be represented with the inverted spherical representation $\bm{p'} = [\theta, \phi, w']^\top$ as:
\begin{equation}
\begin{cases}
\theta = \arctan\left(\frac{y}{x}\right), \\
\phi = \arcsin\left(\frac{z}{\|r\|}\right), \\
 w' = \frac{1}{r},
\end{cases}
\end{equation}
where $w'$ is the inverted depth and $r = \sqrt{x^2 + y^2 + z^2} > 1$. The inverted spherical representation can be converted to Cartesian by:
\begin{equation}
\begin{cases}
x = \frac{\sin{\phi}\cos{\theta}}{w'}, \\
y = \frac{\sin{\phi}\sin{\theta}}{w'}, \\
z = \frac{\cos{\phi}}{w'}.
\end{cases}
\end{equation}

\paragraph{Comparison Results.}
The quantitative results presented in \Tref{tab:spherical} and \Tref{tab:spherical_near_far} show that homogeneous representation consistently outperforms inverted spherical representation in indoor and near-object scenarios. Although inverted depth $w'$ effectively represents distant points in its range \((0, 1]\), its mapping of points with depth within $1$ to its range \([1, +\infty)\) hinders the performance on nearby objects. In contrast, homogeneous coordinates offer a balanced representation of near and far objects, using the weight $w$ to account for depths.

\begin{table*}[t!]
  \small
\resizebox{1.0\linewidth}{!}{
  \tabcolsep=0.07cm
  \begin{tabular}{@{\hspace{0mm}}c@{\hspace{7mm}}c@{\hspace{3mm}}|@{\hspace{3mm}}c@{\hspace{3mm}}c@{\hspace{3mm}}c@{\hspace{3mm}}|@{\hspace{3mm}}c@{\hspace{3mm}}c@{\hspace{3mm}}c@{\hspace{3mm}}|@{\hspace{3mm}}c@{\hspace{3mm}}c@{\hspace{3mm}}c@{\hspace{5mm}}}
  \toprule
   & Dataset & \multicolumn{3}{c|@{\hspace{3mm}}}{Mip-NeRF 360 Dataset}  & \multicolumn{3}{c|@{\hspace{3mm}}}{Tanks\&Temples} & \multicolumn{3}{c@{\hspace{5mm}}}{DL3DV-10K Benchmark}\\
   & Method \textbar{} Metric
    & SSIM$^\uparrow$   & PSNR$^\uparrow$    & LPIPS$^\downarrow$
    & SSIM$^\uparrow$   & PSNR$^\uparrow$    & LPIPS$^\downarrow$ 
    & SSIM$^\uparrow$   & PSNR$^\uparrow$    & LPIPS$^\downarrow$\\
    \midrule

\multirow{2}{*}
   & Inverted Spherical &   \textbf{0.828}         & \textbf{27.93}       &  0.196   & 0.855 & 24.15 &     0.169   &       0.918 &       29.85 &       0.117\\
   & \HGS (Ours) &    \textbf{0.828} & 27.92  & \textbf{0.194} & \textbf{0.858} & \textbf{24.27} & \textbf{0.166} & \textbf{0.919} & \textbf{29.93} & \textbf{0.114} \\
  \bottomrule
  \end{tabular}
  }\vspace{-2mm}
  \caption{
    \textbf{Inverted spherical vs.~homogeneous representations.} This table reports the performance of Inverted Spherical and Homogeneous (\HGS) methods on three datasets: Mip-NeRF 360~\cite{mipnerf360}, Tanks\&Temples~\cite{tanks}, and DL3DV-10K Benchmark~\cite{dl3dv}. Results are evaluated using three metrics: SSIM, PSNR, and LPIPS. \HGS consistently shows competitive or better performance across all datasets. 
  }\vspace{-2mm}
  \label{tab:spherical}
\end{table*}

\begin{table}[tp]
\centering
\resizebox{1.0\linewidth}{!}{
\begin{tabular}{c|ccc|ccc}
\toprule
\multirow{2}{*}{Method} & \multicolumn{3}{c|}{Tanks\&Temples (\textbf{Near})} & \multicolumn{3}{c}{Tanks\&Temples (\textbf{Far})} \\ 
\cline{2-7} 
& SSIM$^\uparrow$ & PSNR$^\uparrow$ & LPIPS$^\downarrow$ & SSIM$^\uparrow$ & PSNR$^\uparrow$ & LPIPS$^\downarrow$ \\ 
\midrule
 Inverted Spherical & 0.882 & 25.54 & 0.137 & \textbf{0.976} & \textbf{30.78} & 0.032 \\
 \HGS (Ours) & \textbf{0.883} & \textbf{25.69} & \textbf{0.135} & \textbf{0.976} & 30.77 & \textbf{0.031} \\ 
\midrule
\midrule
\multirow{2}{*}{Method} & \multicolumn{3}{c|}{DL3DV-10K (\textbf{Near})} & \multicolumn{3}{c}{DL3DV-10K (\textbf{Far})} \\ 
\cline{2-7} 
& SSIM$^\uparrow$ & PSNR$^\uparrow$ & LPIPS$^\downarrow$ & SSIM$^\uparrow$ & PSNR$^\uparrow$ & LPIPS$^\downarrow$ \\ 
\midrule
 Inverted Spherical & 0.926 & 30.46 & 0.101 & \textbf{0.993} & 40.43 & \textbf{0.013} \\
 \HGS (Ours) & \textbf{0.927} & \textbf{30.53} & \textbf{0.098} & \textbf{0.993} & \textbf{40.55} & \textbf{0.013} \\ 
\midrule
\midrule
\multirow{2}{*}{Method} & \multicolumn{3}{c|}{Mip-NeRF360 (\textbf{Indoor})} & \multicolumn{3}{c}{Mip-NeRF360 (\textbf{Outdoor})} \\ 
\cline{2-7} 
 & SSIM$^\uparrow$ & PSNR$^\uparrow$ & LPIPS$^\downarrow$ & SSIM$^\uparrow$ & PSNR$^\uparrow$ & LPIPS$^\downarrow$ \\ 
\midrule
Inverted Spherical & 0.917 & 30.40 & \textbf{0.157} & 0.746 & 25.13 & 0.223 \\
\HGS (Ours) & \textbf{0.930} & \textbf{31.38} & 0.163 & \textbf{0.747} & \textbf{25.14} & \textbf{0.219} \\
\bottomrule
\end{tabular}
}\vspace{-2mm}
\caption{\textbf{Performance of inverted spherical and homogeneous methods across different scenarios.} 
This table highlights the performance of Homogeneous and Spherical representations on Tanks\&Temples, DL3DV-10K, and Mip-NeRF360 datasets. Homogeneous representation consistently outperforms the inverted Spherical representation in indoor and near-object scenarios.}\vspace{-2mm}
\label{tab:spherical_near_far}
\end{table}

\section{More Convergence Analysis}
\label{supp:convergence}
While the main paper presents a convergence analysis in a simple setup, we further analyze the convergence behaviors.

\begin{figure}[t]
  \centering
  \includegraphics[width=\linewidth]{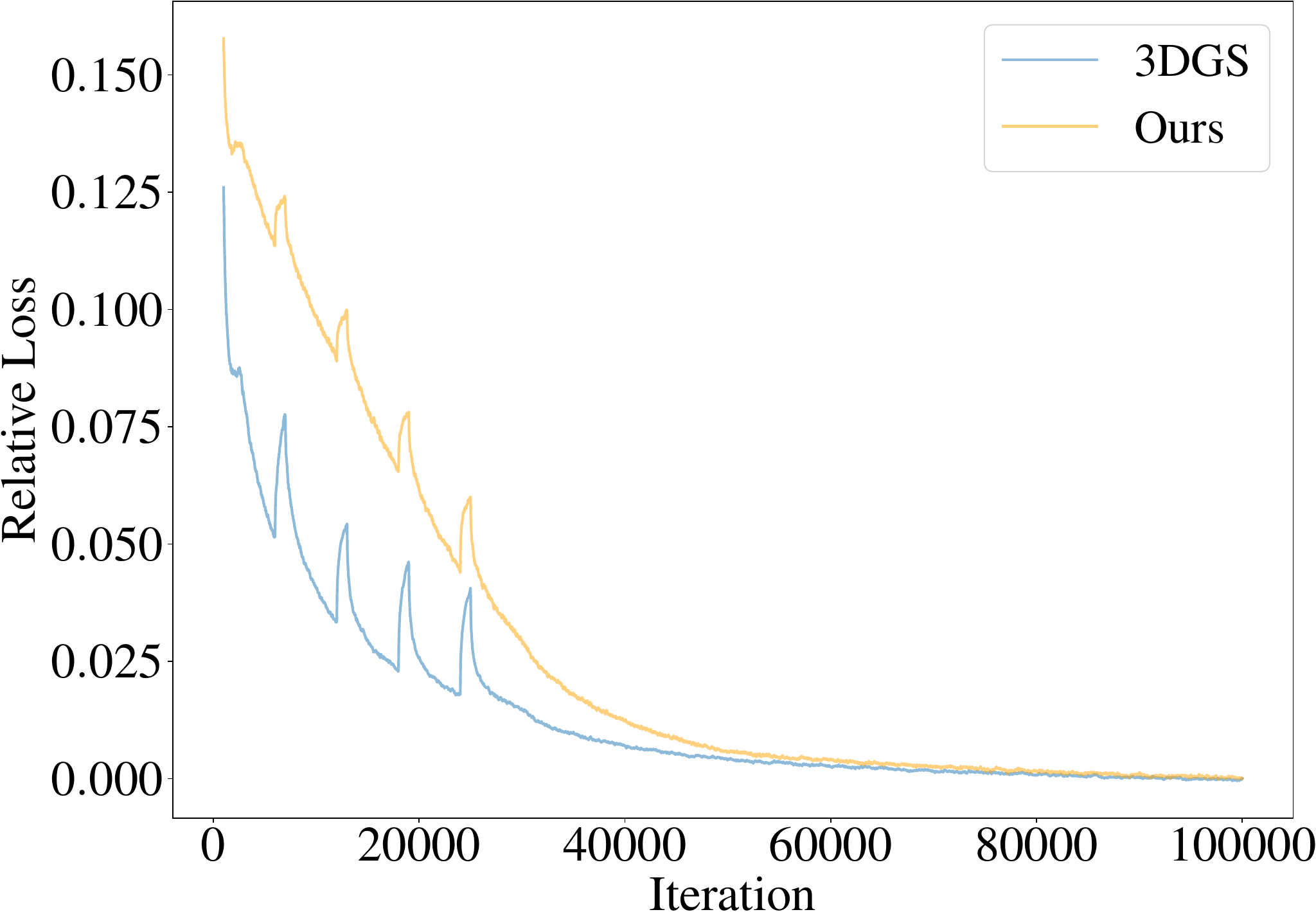}\vspace{-2mm}
   \caption{
   \textbf{Loss convergence of 3DGS and HoGS.}
   The loss curves indicate that with our setup on unbounded scenes, both 3DGS and HoGS have not fully converged by $30,000$ iterations, while with $50,000$ iterations they become nearly minimal.
}  \vspace{-2mm}
   \label{fig:supp_loss_compare}
\end{figure}
\begin{figure}[t]
  \centering
  \resizebox{1.00\linewidth}{!}{
  \begin{tabular}{cc}
    \includegraphics[width=0.325\textwidth]{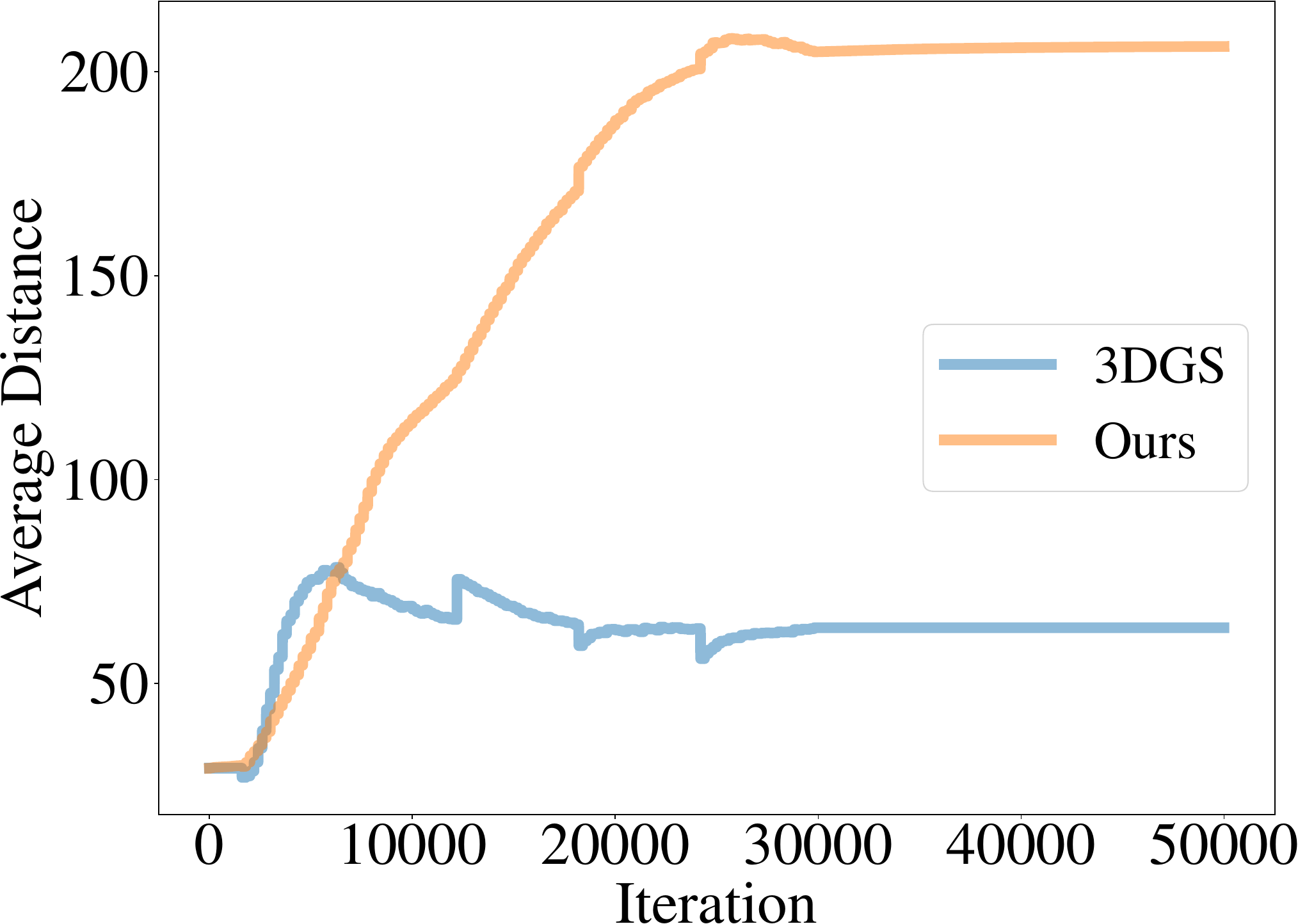} &
    \includegraphics[width=0.325\textwidth]{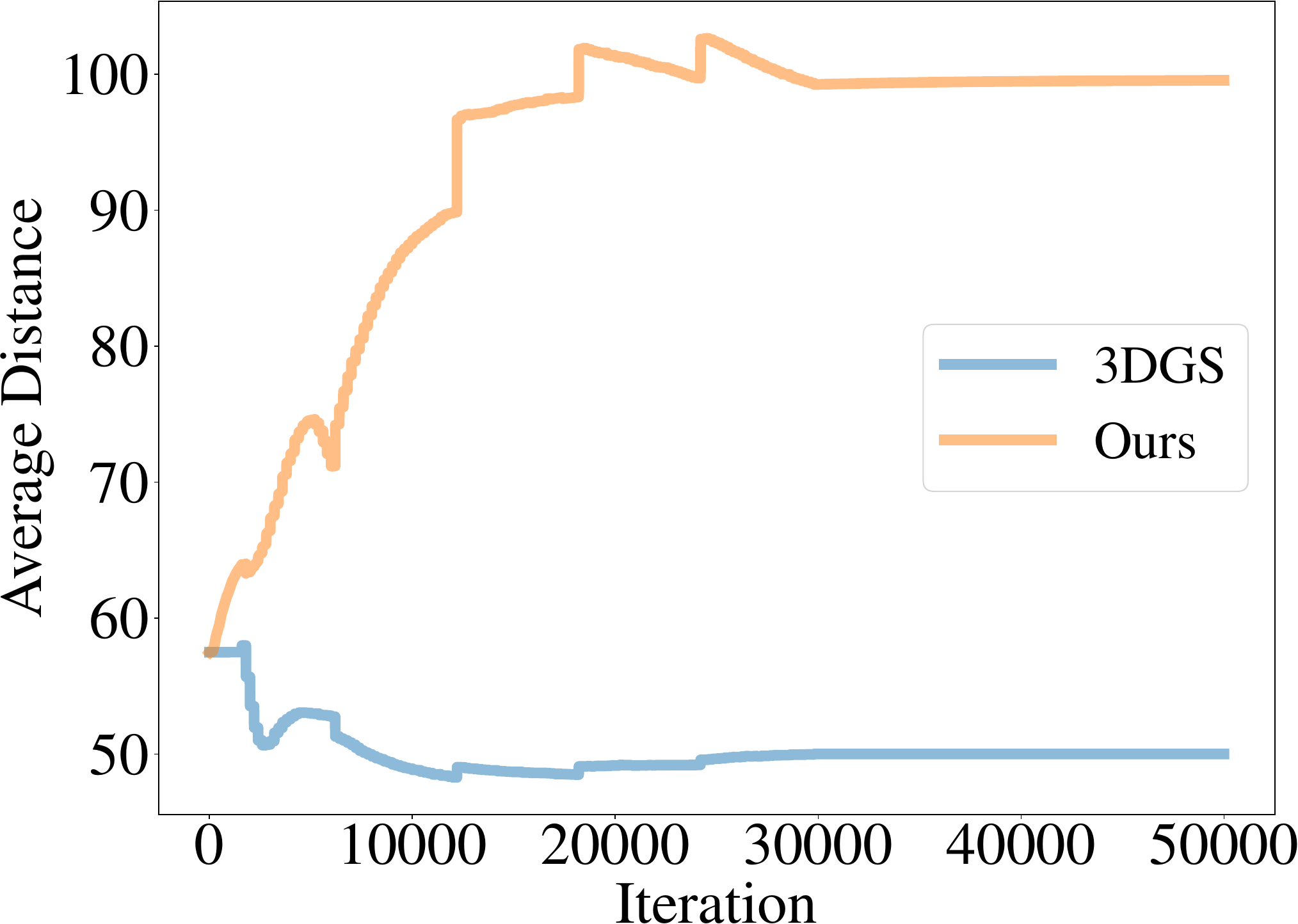} \\
    \multicolumn{1}{c}{(a) TRAIN} & \multicolumn{1}{c}{(b) DL3 21}
  \end{tabular}\vspace{-2mm}
}
   \caption{
   \textbf{Average distance of $10\%$ farthest points.} In both the TRAIN and DL3DV 21 scenes, the $10\%$ farthest points in our method reach significantly farther distances compared to 3DGS. By the end of the training, the farthest points in our method reach approximately $1,200$ meters in the TRAIN dataset (compared to $410$ meters in 3DGS) and $520$ meters in DL3 Scene 21 (compared to $97$ meters in 3DGS) in the physical space.
}  \vspace{-2mm}
   \label{fig:supp_farthest_points}
\end{figure}

\paragraph{Extended Training.}
The main paper uses $50,000$ iterations for both the original 3DGS and our HoGS. To validate the number of iterations for unbounded scenes, we present the training process of $100,000$ iterations in an unbounded scene (the TRAIN scene in the Tank\&Temples dataset~\cite{tanks}) in \fref{fig:supp_loss_compare}. The loss curves in \fref{fig:supp_loss_compare} show the relative loss, where the loss values are normalized by setting the loss at $50,000$ iterations as the reference point. Both 3DGS and our method reduce the training losses until around $50,000$ iterations, which are almost converging there, demonstrating that $50,000$ iterations are reasonable for unbounded scene reconstruction in the main paper's experiments.

\paragraph{Convergence Behavior in Real Scenes.}
Here, we analyze the convergence behaviors in real scenes by tracking the average displacement of the Gaussian centers farthest $10\%$ from the world origin over iterations. 
\Fref{fig:supp_farthest_points} show that far points in our method move more quickly to distant positions than the original 3DGS in unbounded scenes (TRAIN~\cite{tanks} and DL3DV Scene 21~\cite{dl3dv}), underscoring the efficiency of our approach in reconstructing distant objects.

\begin{figure}[htbp]
  \centering
  \resizebox{1.00\linewidth}{!}{
  \begin{tabular}{cc}
    \includegraphics[width=0.425\textwidth]{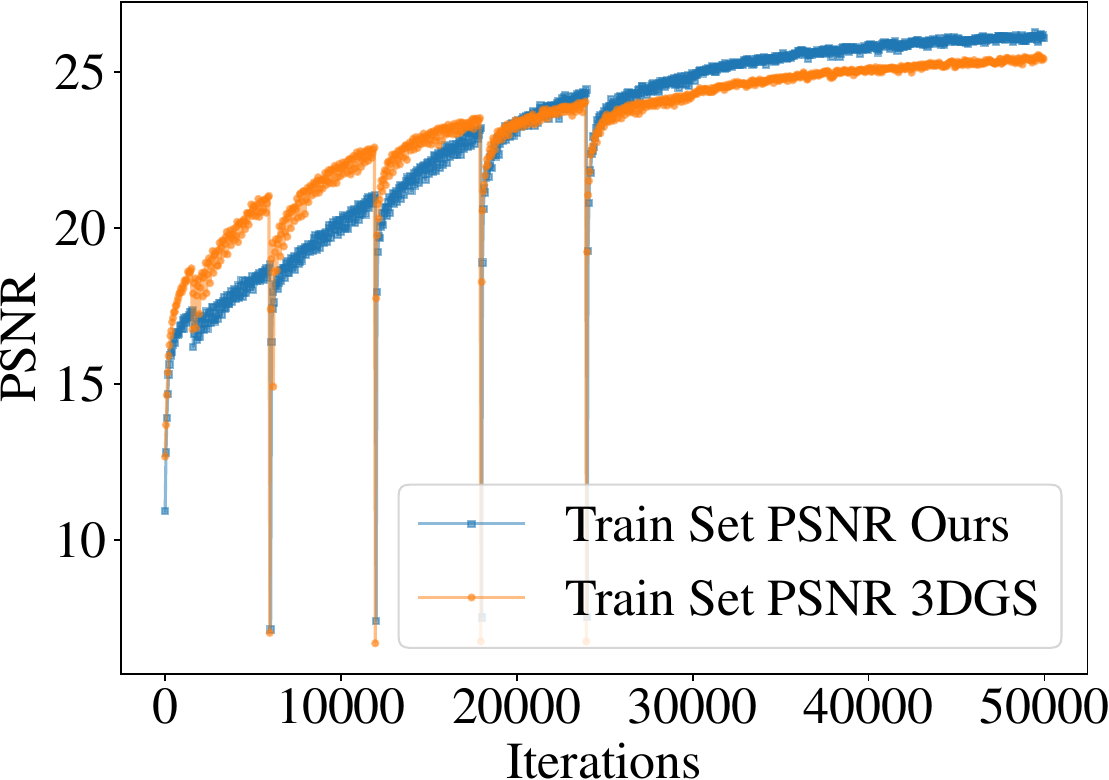} &
    \includegraphics[width=0.425\textwidth]{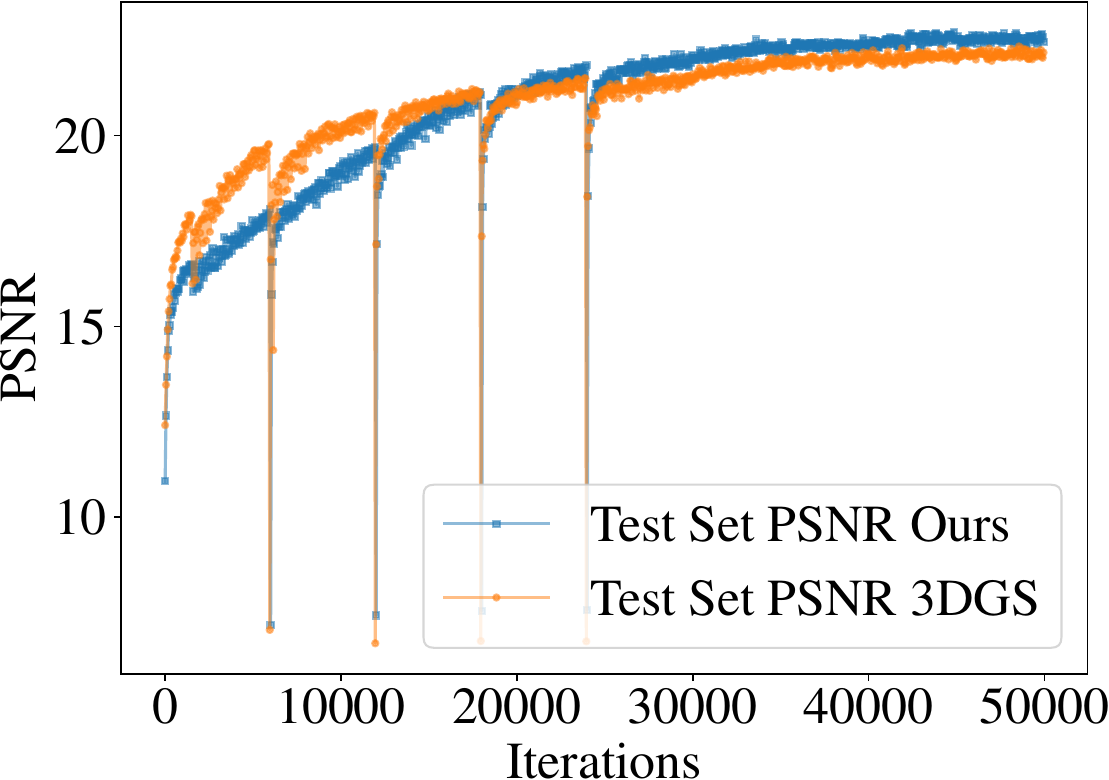} \\
    \multicolumn{1}{c}{(a) Train PSNR curve} & \multicolumn{1}{c}{(b) Test PSNR curve}
  \end{tabular}\vspace{-2mm}
}
   \caption{
   PSNR-iteration curve of our HoGS and 3DGS. HoGS exhibits a slower initial convergence due to the homogeneous optimization of near and far objects, requiring Gaussians to propagate across the scene. However, as training progresses, HoGS surpasses 3DGS in PSNR, demonstrating its superior reconstruction fidelity and global consistency.
}  \vspace{-2mm}
   \label{fig:supp_curve}
\end{figure}

\paragraph{PSNR Convergence Analysis.}
\label{supp:psnr_curve}
We also show a PSNR-iteration curve comparing HoGS and 3DGS as shown in \fref{fig:supp_curve}. HoGS shows a slower initial convergence due to the homogeneous optimization of near and far objects, which requires Gaussians to propagate across the entire scene. However, as training progresses, HoGS surpasses 3DGS in PSNR, demonstrating superior reconstruction fidelity and global consistency.

\section{More Design Choices}
\label{supp:design}

\subsection{Skybox Initialization}
\label{supp:skibox}

Following the methods in \cite{hierarchical, gaustudio}, we can essentially use skybox~\cite{hierarchical} initialization to enhance the points in distant scenes and the 3DGS and our proposed methods. 
We thus test the skybox initialization to both the 3DGS and our proposed methods. Specifically, we added $100,000$ blue points as initial points on the upper hemisphere, with a radius of $1,000$ unit distance, which corresponds to approximately $1.1$ kilometers in DL3DV Scene 21 (estimated using the base width of the George Washington statue in front of the Indiana Statehouse) and $2.1$ kilometers in the TRAIN dataset (estimated from the width of railway tracks). As shown in \Tref{tab:supp_skybox}, the results indicate that incorporating a skybox improves the accuracy of distant scenes in the 3DGS method. This enhancement arises because the skybox provides additional guidance for representing distant elements like clouds, otherwise challenging for 3DGS to capture.

In contrast, adding a skybox to our method does not improve the accuracy of distant scenes. This is because our approach, even without skybox initialization, effectively leverages the attributes of homogeneous coordinates to represent both near and distant objects equally. Our method accurately represents distant areas without additional priors by inherently keeping consistent scaling across varying depths.

\begin{table}[tp]
\centering
\resizebox{1.0\linewidth}{!}{
\begin{tabular}{c|ccc|ccc}
\toprule
\multirow{2}{*}{Method} & \multicolumn{3}{c|}{TRAIN (\textbf{Near})} & \multicolumn{3}{c}{TRAIN (\textbf{Far})} \\ 
\cline{2-7} 
& SSIM$^\uparrow$ & PSNR$^\uparrow$ & LPIPS$^\downarrow$ & SSIM$^\uparrow$ & PSNR$^\uparrow$ & LPIPS$^\downarrow$ \\ 
\midrule
 3DGS w/o skybox & \textbf{0.854} & \textbf{23.86} & 0.151 & 0.960 & 28.49 & 0.063 \\
 3DGS w/ skybox & 0.851 & 23.67 & 0.153 & 0.962 & 28.58 & 0.063 \\
 Ours w/o skybox & \textbf{0.854} & 23.70 & \textbf{0.147} & \textbf{0.976} & \textbf{30.42} & \textbf{0.044} \\ 
 Ours w/ skybox & 0.849 & 23.33 & 0.149 & 0.975 & 30.23 & 0.046 \\ 
\midrule
\midrule
\multirow{2}{*}{Method} & \multicolumn{3}{c|}{DL3DV Scene 21 (\textbf{Near})} & \multicolumn{3}{c}{DL3DV Scene 21 (\textbf{Far})} \\ 
\cline{2-7} 
& SSIM$^\uparrow$ & PSNR$^\uparrow$ & LPIPS$^\downarrow$ & SSIM$^\uparrow$ & PSNR$^\uparrow$ & LPIPS$^\downarrow$ \\ 
\midrule
 3DGS w/o skybox & 0.870 & 27.72 & 0.147 & 0.972 & 31.23 & 0.042 \\
 3DGS w/ skybox & 0.867 & 27.78 & 0.152 & 0.976 & 32.16 & 0.038 \\
 Ours w/o skybox & 0.889 & 28.81 & 0.122 & \textbf{0.986} & \textbf{35.48} & \textbf{0.022} \\ 
 Ours w/ skybox & \textbf{0.890} & \textbf{28.83} & \textbf{0.120} & \textbf{0.986} & 35.15 & \textbf{0.022} \\ 
\bottomrule
\end{tabular}
}\vspace{-2mm}
\caption{\textbf{Impact of skybox on near and far scenarios.} 
This table compares the performance of 3DGS and our method (w/ and w/o skybox) on TRAIN and DL3DV Scene 21 datasets across Near and Far scenarios. Incorporating a skybox initialization can slightly improve the accuracy of distant scenes in the original 3DGS method but not in our method. Our method effectively represents both near and far objects even without skybox initialization.}\vspace{-2mm}
\label{tab:supp_skybox}
\end{table}

\begin{figure*}[htbp]
  \centering
  \includegraphics[width=\textwidth]{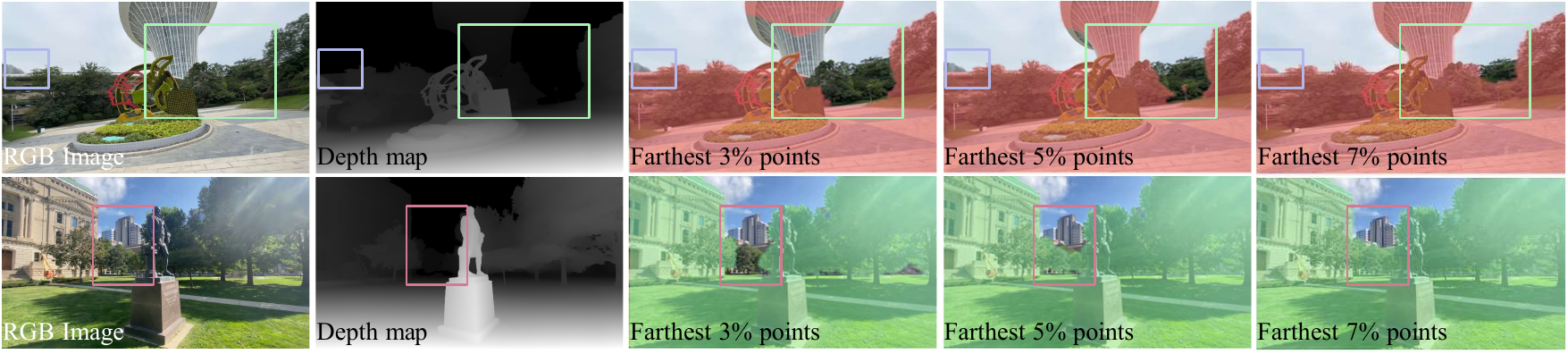}\vspace{-2mm}
   \caption{
   \textbf{Impact of threshold selection on near and far object classification.}
   We evaluate the effects of changing the threshold used to classify far and near objects in the depth mask used for the experiment. When the threshold is reduced to $3\%$, near objects such as parts of buildings and shrubs (green bounding box) and park trees (red bounding box) are mistakenly classified as far. Conversely, increasing the threshold to $7\%$ results in some far objects, like distant buildings, being misclassified as near, as shown by the blue bounding box. To balance accuracy and meaningful far-object coverage, we select the $5~\%$ threshold as the optimal trade-off. }\vspace{-2mm}
   \label{fig:supp_depth_threshold}
\end{figure*}
\begin{table}[tp]
\centering
\resizebox{1.0\linewidth}{!}{
\begin{tabular}{c|c|ccc|ccc}
\toprule
\multirow{2}{*}{Scene} & \multirow{2}{*}{Threshold} & \multicolumn{3}{c|}{\textbf{Near}} & \multicolumn{3}{c}{\textbf{Far}} \\ 
\cline{3-8} 
& & SSIM$^\uparrow$ & PSNR$^\uparrow$ & LPIPS$^\downarrow$ & SSIM$^\uparrow$ & PSNR$^\uparrow$ & LPIPS$^\downarrow$ \\ 
\midrule
\multirow{3}{*}{TRAIN} & $7~\%$ & 0.858 & 23.82 & 0.143 & 0.972 & 29.94 & 0.048 \\
       & $5~\%$ & 0.854 & 23.70 & 0.147 & 0.976 & 30.43 & 0.044 \\
       & $3~\%$ & 0.848 & 23.53 & 0.151 & 0.982 & 31.25 & 0.039 \\ 
 \midrule \midrule
 \multirow{3}{*}{DL3DV Scene 21} & $7~\%$ & 0.893 & 29.08 & 0.118 & 0.982 & 34.37 & 0.025 \\
       & $5~\%$ & 0.889 & 28.81 & 0.122 & 0.986 & 35.48 & 0.022 \\
       & $3~\%$ & 0.885 & 28.60 & 0.125 & 0.989 & 36.80 & 0.019 \\ 
 \midrule \midrule
 \multirow{3}{*}{DL3DV Scene 97} & $7~\%$ & 0.965 & 34.63 & 0.047 & 0.989 & 38.69 & 0.043 \\
       & $5~\%$ & 0.962 & 34.12 & 0.051 & 0.992 & 40.41 & 0.039 \\
       & $3~\%$ & 0.959 & 33.72 & 0.054 & 0.995 & 42.48 & 0.036 \\
\bottomrule
\end{tabular}
}\vspace{-2mm}
\caption{\textbf{Impact of depth threshold selection on near and far object accuracy.} 
This table evaluates the performance of near and far objects across different depth thresholds ($7~\%$, $5~\%$, and $3~\%$) on the TRAIN, DL3DV Scene 21, and DL3DV Scene 97 datasets. As the threshold decreases, the accuracy for far objects consistently improves (\eg, PSNR increases from $29.94$ to $31.25$ on TRAIN and from $34.37$ to $36.80$ on DL3DV Scene 21), driven by a greater share of sky pixels characterized by uniform colors and high accuracy. However, this improvement in the metrics does not take what we really want to evaluate, \eg, buildings and mountains, into account.}\vspace{-2mm}
\label{tab:depth_threshold}
\end{table}

\begin{figure*}[tp]
  \centering
  \includegraphics[width=\textwidth]{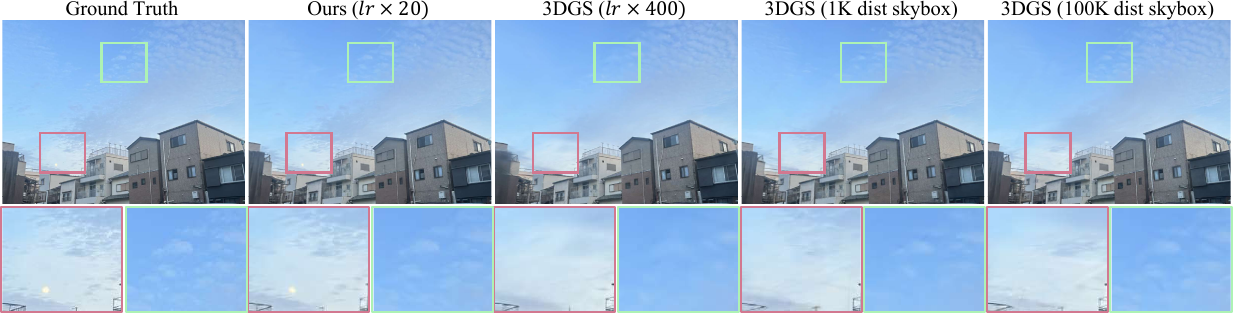}\vspace{-2mm}
   \caption{
   \textbf{Reconstruction of an infinitely far (\ie, the Moon) scene.}
     With a 20x increase in the learning rate (\(lr\)), our method successfully represents the Moon, even at its (near-)infinity distances, as highlighted by the red bounding box. In contrast, 3DGS fails to achieve similar results, even when the \(lr\) is largely increased, or when incorporating a skybox with different ($1,000$ and $100,000$) unit distances. Notably, while adding a skybox to 3DGS improves the fidelity of cloud details compared to simply increasing \(lr\), it remains incapable of generating a realistic appearance of the Moon.}\vspace{-2mm}
   \label{fig:supp_moon}
\end{figure*}

\begin{table*}[tp]
\centering
\resizebox{1.0\linewidth}{!}{
\begin{tabular}{c|ccc|ccc|ccc|ccc|ccc}
\toprule
\multirow{2}{*}{Method} & \multicolumn{3}{c|}{\textbf{BICYCLE}} & \multicolumn{3}{c|}{\textbf{FLOWERS}} & \multicolumn{3}{c|}{\textbf{GARDEN}}  & \multicolumn{3}{c|}{\textbf{STUMP}}  & \multicolumn{3}{c}{\textbf{TREEHILL}} \\ 
\cline{2-16} 
& SSIM$^\uparrow$ & PSNR$^\uparrow$ & LPIPS$^\downarrow$ & SSIM$^\uparrow$ & PSNR$^\uparrow$ & LPIPS$^\downarrow$ & SSIM$^\uparrow$ & PSNR$^\uparrow$ & LPIPS$^\downarrow$ & SSIM$^\uparrow$ & PSNR$^\uparrow$ & LPIPS$^\downarrow$ & SSIM$^\uparrow$ & PSNR$^\uparrow$ & LPIPS$^\downarrow$ \\ 
\midrule
 3DGS & 0.770 & 25.59 & 0.219 & 0.609 & 21.36 & 0.347 & 0.869 & 27.68 & 0.110 & 0.775 & 26.66 & 0.223 & 0.649 & 22.63 & 0.332 \\
 Scaffold-GS & 0.768 & 25.57 & 0.228 & 0.610 & 21.67 & 0.337 & 0.869 & \textbf{27.92} & 0.111 & 0.777 & 26.88 & 0.228 & \textbf{0.664} & \textbf{23.53} & 0.312 \\ 
 \HGS (Ours) & \textbf{0.790} & \textbf{25.84} & \textbf{0.183} & \textbf{0.638} & \textbf{22.18} & \textbf{0.311} & \textbf{0.874} & 27.83 & \textbf{0.098} & \textbf{0.785} & \textbf{26.92} & \textbf{0.195} & 0.649 & 22.95 & \textbf{0.310} \\
\midrule \midrule
\multirow{2}{*}{Method} & \multicolumn{3}{c|}{\textbf{BONSAI}} & \multicolumn{3}{c|}{\textbf{COUNTER}} & \multicolumn{3}{c|}{\textbf{KITCHEN}}  & \multicolumn{3}{c|}{\textbf{ROOM}} \\ 
\cline{2-13} 
& SSIM$^\uparrow$ & PSNR$^\uparrow$ & LPIPS$^\downarrow$ & SSIM$^\uparrow$ & PSNR$^\uparrow$ & LPIPS$^\downarrow$ & SSIM$^\uparrow$ & PSNR$^\uparrow$ & LPIPS$^\downarrow$ & SSIM$^\uparrow$ & PSNR$^\uparrow$ & LPIPS$^\downarrow$  \\ 
\midrule
 3DGS & 0.946 & 32.28 & 0.178 & 0.914 & 29.33 & 0.182 & 0.931 & 31.41 & 0.115 & 0.925 & 31.93 & 0.198  \\
 Scaffold-GS & \textbf{0.950} & \textbf{33.08} & 0.169 & \textbf{0.919} & \textbf{29.81} & \textbf{0.177} & \textbf{0.934} & 31.94 & \textbf{0.113} & \textbf{0.932} & \textbf{32.34} & \textbf{0.181}  \\ 
 \HGS (Ours) & 0.948 & 32.56 & \textbf{0.164} & 0.913 & 29.25 & 0.179 & 0.932 & \textbf{32.04} & 0.114 & 0.925 & 31.67 & 0.196  \\
 \bottomrule
\end{tabular}
}\vspace{-2mm}
\caption{
\textbf{Comparisons for Mip-NeRF360 outdoor and indoor scenes.} 3DGS-based methods' SSIM, PSNR and LPIPS scores for Mip-NeRF360 scenes. Outdoor scenes are listed above, while indoor scenes are listed below.
}\vspace{-2mm}
\label{tab:360scenes}
\end{table*}
\begin{table*}[t!]
  \small
\resizebox{1.0\linewidth}{!}{
  \tabcolsep=0.07cm
  \begin{tabular}{@{\hspace{2mm}}c@{\hspace{5mm}}c@{\hspace{3mm}}|@{\hspace{3mm}}c@{\hspace{3mm}}c@{\hspace{3mm}}c@{\hspace{3mm}}|@{\hspace{3mm}}c@{\hspace{3mm}}c@{\hspace{3mm}}c@{\hspace{3mm}}|@{\hspace{3mm}}c@{\hspace{3mm}}c@{\hspace{3mm}}c@{\hspace{2mm}}}
  \toprule
   & Dataset & \multicolumn{3}{c|@{\hspace{3mm}}}{Mip-NeRF 360 Dataset}  & \multicolumn{3}{c|@{\hspace{3mm}}}{Tanks\&Temples} & \multicolumn{3}{c@{\hspace{5mm}}}{DL3DV-10K Benchmark}\\
   & Method{\hspace{3mm}} / {\hspace{3mm}}Metric
    & SSIM$^\uparrow$   & PSNR$^\uparrow$    & LPIPS$^\downarrow$
    & SSIM$^\uparrow$   & PSNR$^\uparrow$    & LPIPS$^\downarrow$ 
    & SSIM$^\uparrow$   & PSNR$^\uparrow$    & LPIPS$^\downarrow$\\
    \midrule
    & Mip-Splatting~\cite{Yu2024MipSplatting} & 0.835 & 27.97 & 0.182 & 0.856 & 24.01 & 0.160 & 0.917 & 29.44 & 0.116 \\ 
    & Multi-Scale 3DGS~\cite{yan2024multiscale3dgaussiansplatting} & 0.821 & 27.73 & 0.211 & 0.846 & 23.92 & 0.180 & 0.893 & 28.23 & 0.148 \\ 
   & \HGS (Ours) &    0.828 & 27.92 & 0.194 & 0.858 & 24.27 & 0.166 & 0.919 & 29.93 & 0.114 \\
  \bottomrule
  \end{tabular}
  }\vspace{-2mm}
  \caption{
    \textbf{Comparison with Mip-Splatting and Multi-Scale 3D Gaussian Splatting.}
  }\vspace{-2mm}
  \label{tab:comparison_clab}
\end{table*}

\subsection{Threshold for Near vs.~Far in Experiment}
\label{supp:threshold}

In the main paper's experiment, to evaluate the accuracy for near and far objects, we use Depth Anything V2~\cite{depthanything} to generate depth maps from input images at their original resolution. We define distant areas as these maps' farthest $5~\%$ of depth values. This threshold is effective because adopting the $0~\%$ threshold would misclassify distant mountains and buildings as near objects, evaluating far objects' accuracy only on limited sky regions. Here, we additionally tested thresholds of $3~\%$ and $7~\%$ as shown in \fref{fig:supp_depth_threshold} and \Tref{tab:depth_threshold}. As the threshold changes from $7~\%$ to $5~\%$ to $3~\%$, the accuracy for far objects improves. This is because the number of pixels classified as distant objects decreases while the number of pixels representing the sky remains constant. Typically characterized by uniform colors and high accuracy, the sky becomes a more significant proportion of the far-object pixels as the threshold is reduced. Consequently, the overall accuracy improves with an increasing share of sky pixels in the far-object category.

However, it is essential to note that while reducing the threshold improves the accuracy metrics, it does not measure the representation of other \emph{distant} objects, such as mountains or buildings, which are critical for photorealistic scene reconstruction (see \fref{fig:supp_depth_threshold}). Hence, we set a $5~\%$ threshold in the main paper's experiment, which provides a balanced trade-off between including enough distant objects for meaningful evaluation and avoiding the misclassification of nearby elements. 

\section{Representing Infinitely Far Objects}
\label{supp:moon}

We collected a custom dataset containing scenes with the Moon to assess the reconstruction capability of objects at \emph{infinitely} far away\footnote{Due to its proprietary nature, this dataset is used exclusively for experimental purposes.}. \Fref{fig:supp_moon} shows a visual comparison. With our method, increasing the learning rate (\(lr\)) for the weight parameter \(w\) by 20x enabled the accurate reconstruction of the Moon. In contrast, 3DGS failed to reconstruct the Moon even with a large \(lr\), with an increase of 400x, producing artifacts that disappear when the viewpoint changes. Adding a skybox with a $1,000$ unit or $100,000$ unit distance to 3DGS also failed to produce a realistic appearance of the Moon, where these distances are approximately $500$~m and $50,000$~m, respectively, in the physical space.

\begin{figure}[t]
  \centering
  \includegraphics[width=\linewidth]{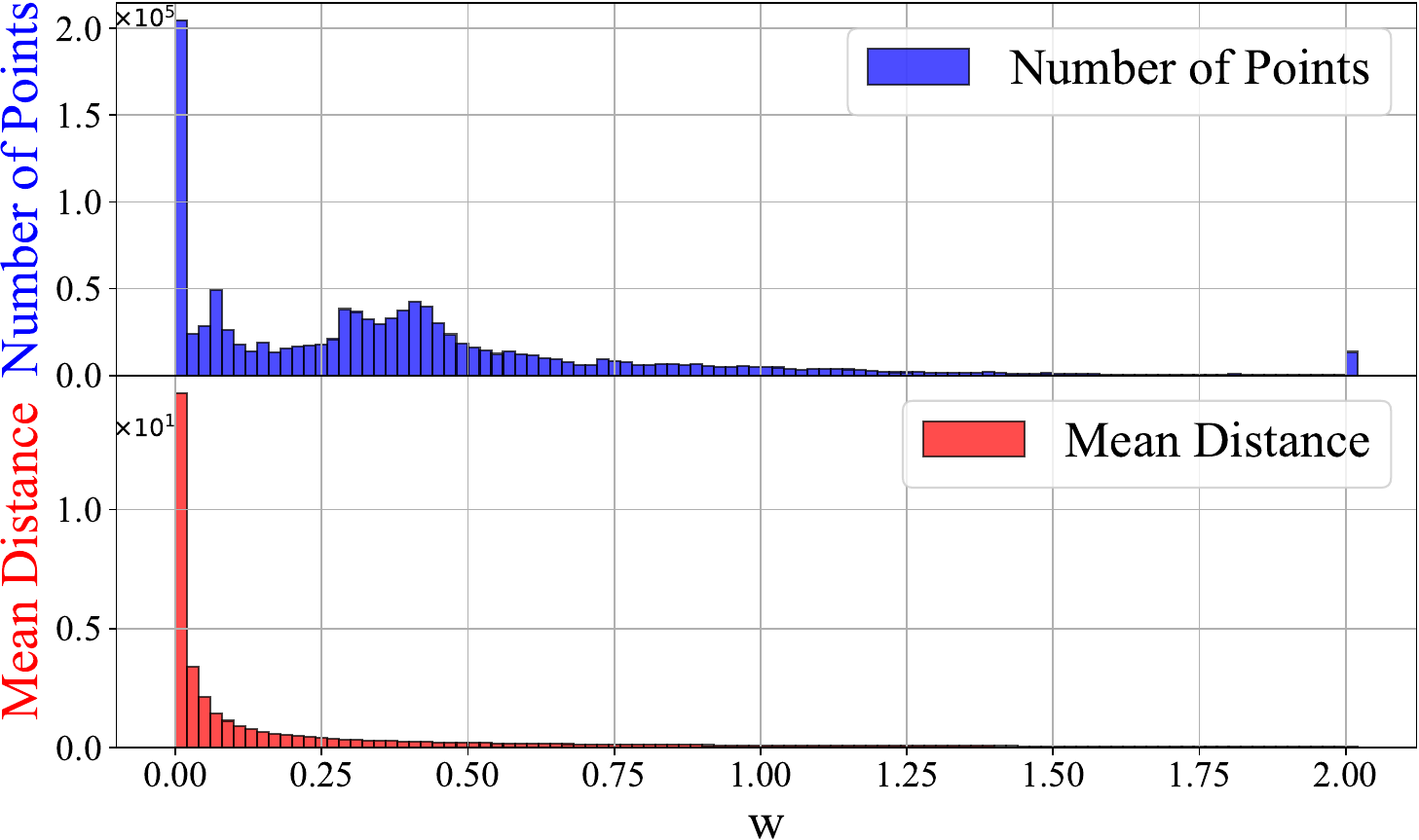}\vspace{-2mm}
\caption{Histogram of $w$.} 
   \label{fig:ablation1}
\end{figure}


\begin{table*}[tp]
\centering
\resizebox{1.0\linewidth}{!}{
\begin{tabular}{c|c|ccc|ccc|ccc}
\toprule
\multirow{2}{*}{Scene} & \multirow{2}{*}{Method} & \multicolumn{3}{c|}{\textbf{Overall}} & \multicolumn{3}{c|}{\textbf{Near}} & \multicolumn{3}{c}{\textbf{Far}} \\ 
\cline{3-11} 
& & SSIM$^\uparrow$ & PSNR$^\uparrow$ & LPIPS$^\downarrow$ & SSIM$^\uparrow$ & PSNR$^\uparrow$ & LPIPS$^\downarrow$ & SSIM$^\uparrow$ & PSNR$^\uparrow$ & LPIPS$^\downarrow$ \\ 
\midrule
\multirow{3}{*}{TRAIN} & 3DGS & 0.811 & 22.16 & 0.214 & 0.854 & 23.86 & 0.151 & 0.960 & 28.49 & 0.063 \\
       & Scaffold-GS & 0.825 & \textbf{22.69} & 0.200 & \textbf{0.859} & \textbf{23.87} & \textbf{0.146} & 0.969 & 30.36 & 0.053 \\
       & \HGS (Ours) & \textbf{0.828} & 22.60 & \textbf{0.191} & 0.854 & 23.70 & 0.147 & \textbf{0.976} & \textbf{30.43} & \textbf{0.044} \\ 
 \midrule \midrule
 \multirow{3}{*}{TRUCK} & 3DGS & 0.878 & 25.50 & 0.152 & 0.908 & 27.42 & 0.130 & 0.973 & 30.31 & 0.022 \\
       & Scaffold-GS & 0.883 & 25.92 & \textbf{0.139} & 0.911 & \textbf{27.87} & \textbf{0.118} & 0.974 & 30.71 & 0.020  \\
       & \HGS (Ours) & \textbf{0.887} & \textbf{25.94} & 0.141 & \textbf{0.913} & 27.68 & 0.123 & \textbf{0.977} & \textbf{31.11} & \textbf{0.018} \\ 
\bottomrule
\end{tabular}
}\vspace{-2mm}
\caption{\textbf{Comparisons for Tanks\&Temples scenes with near/far metrics.} 3DGS-based methods' SSIM, PSNR and LPIPS scores for Tanks\&Temples scenes.}\vspace{-2mm}
\label{tab:tandt}
\end{table*}
\begin{table*}[tp]
\centering
\resizebox{1.0\linewidth}{!}{
\begin{tabular}{c|c|ccc|ccc|ccc}
\toprule
\multirow{2}{*}{Scene} & \multirow{2}{*}{Method} & \multicolumn{3}{c|}{\textbf{Overall}} & \multicolumn{3}{c|}{\textbf{Near}} & \multicolumn{3}{c}{\textbf{Far}} \\ 
\cline{3-11} 
& & SSIM$^\uparrow$ & PSNR$^\uparrow$ & LPIPS$^\downarrow$ & SSIM$^\uparrow$ & PSNR$^\uparrow$ & LPIPS$^\downarrow$ & SSIM$^\uparrow$ & PSNR$^\uparrow$ & LPIPS$^\downarrow$ \\ 
\midrule
\multirow{3}{*}{24} & 3DGS & 0.944 & 31.77 & 0.094 & 0.951 & 32.35 & 0.085 & 0.995 & 41.85 & 0.006 \\
       & Scaffold-GS & 0.946 & 31.98 & 0.090 & 0.950 & 32.32 & 0.084 & 0.997 & 44.33 & 0.003 \\
       & \HGS (Ours) & \textbf{0.953} & \textbf{32.70} & \textbf{0.078} & \textbf{0.955} & \textbf{32.95} & \textbf{0.075} & \textbf{0.998} & \textbf{46.82} & \textbf{0.002} \\ 
 \midrule \midrule
 \multirow{3}{*}{26} & 3DGS & 0.904 & 29.48 & 0.185 & 0.911 & 30.13 & 0.176 & 0.994 & 39.67 & 0.007 \\
       & Scaffold-GS & 0.895 & 29.66 & 0.194 & 0.902 & 30.17 & 0.186 & 0.995 & 40.68 & 0.006  \\
       & \HGS (Ours) & \textbf{0.913} & \textbf{30.20} & \textbf{0.165} & \textbf{0.918} & \textbf{30.64} & \textbf{0.159} & \textbf{0.996} & \textbf{41.71} & \textbf{0.005} \\ 
\midrule \midrule
 \multirow{3}{*}{101} & 3DGS & 0.924 & 27.54 & 0.091 & 0.931 & 27.97 & 0.083 & 0.994 & 39.59 & 0.007 \\
       & Scaffold-GS & 0.907 & 26.94 & 0.106 & 0.911 & 27.20 & 0.102 & 0.997 & 41.01 & 0.003  \\
       & \HGS (Ours) & \textbf{0.936} & \textbf{28.26} & \textbf{0.071} & \textbf{0.939} & \textbf{28.50} & \textbf{0.068} & \textbf{0.998} & \textbf{42.40} & \textbf{0.002} \\ 
\bottomrule
\end{tabular}
}\vspace{-2mm}
\caption{\textbf{Comparisons for DL3DV Benchmark indoor scenes.} 3DGS-based methods' SSIM, PSNR and LPIPS scores for DL3DV Benchmark indoor scenes.}\vspace{-2mm}
\label{tab:dl3dv_indoor}
\end{table*}

\begin{table*}[tp]
\centering
\resizebox{1.0\linewidth}{!}{
\begin{tabular}{c|c|ccc|ccc|ccc}
\toprule
\multirow{2}{*}{Scene} & \multirow{2}{*}{Method} & \multicolumn{3}{c|}{\textbf{Overall}} & \multicolumn{3}{c|}{\textbf{Near}} & \multicolumn{3}{c}{\textbf{Far}} \\ 
\cline{3-11} 
& & SSIM$^\uparrow$ & PSNR$^\uparrow$ & LPIPS$^\downarrow$ & SSIM$^\uparrow$ & PSNR$^\uparrow$ & LPIPS$^\downarrow$ & SSIM$^\uparrow$ & PSNR$^\uparrow$ & LPIPS$^\downarrow$ \\ 
\midrule
\multirow{3}{*}{21} & 3DGS & 0.838 & 25.99 & 0.196 & 0.870 & 27.72 & 0.147 & 0.972 & 31.23 & 0.042 \\
       & Scaffold-GS & 0.846 & 27.00 & 0.181 & 0.866 & 28.11 & 0.148 & 0.983 & 34.11 & 0.028 \\
       & \HGS (Ours) & \textbf{0.872} & \textbf{27.80} & \textbf{0.148} & \textbf{0.889} & \textbf{28.81} & \textbf{0.122} & \textbf{0.986} & \textbf{35.48} & \textbf{0.022} \\ 
 \midrule \midrule
 \multirow{3}{*}{69} & 3DGS & 0.858 & 26.41 & 0.165 & 0.891 & 27.77 & 0.131 & 0.972 & 32.83 & 0.027 \\
       & Scaffold-GS & 0.870 & 27.23 & 0.152 & 0.890 & 28.11 & 0.132 & 0.985 & 35.14 & 0.015  \\
       & \HGS (Ours) & \textbf{0.888} & \textbf{27.50} & \textbf{0.129} & \textbf{0.902} & \textbf{28.19} & \textbf{0.115} & \textbf{0.990} & \textbf{36.50} & \textbf{0.010} \\ 
\midrule \midrule
 \multirow{3}{*}{97} & 3DGS & 0.930 & 29.62 & 0.127 & 0.952 & 32.47 & 0.066 & 0.980 & 33.57 & 0.058 \\
       & Scaffold-GS & 0.945 & 31.82 & 0.099 & 0.956 & 32.87 & 0.056 & 0.991 & 38.81 & 0.042  \\
       & \HGS (Ours) & \textbf{0.953} & \textbf{33.12} & \textbf{0.092} & \textbf{0.962} & \textbf{34.12} & \textbf{0.051} & \textbf{0.992} & \textbf{40.41} & \textbf{0.039} \\ 
\bottomrule
\end{tabular}
}\vspace{-2mm}
\caption{\textbf{Comparisons for DL3DV Benchmark outdoor scenes.} 3DGS-based methods' SSIM, PSNR and LPIPS scores for DL3DV Benchmark outdoor scenes.}\vspace{-2mm}
\label{tab:dl3dv_outdoor}
\end{table*}

\begin{figure*}[tp]
  \centering
  \includegraphics[width=\linewidth]{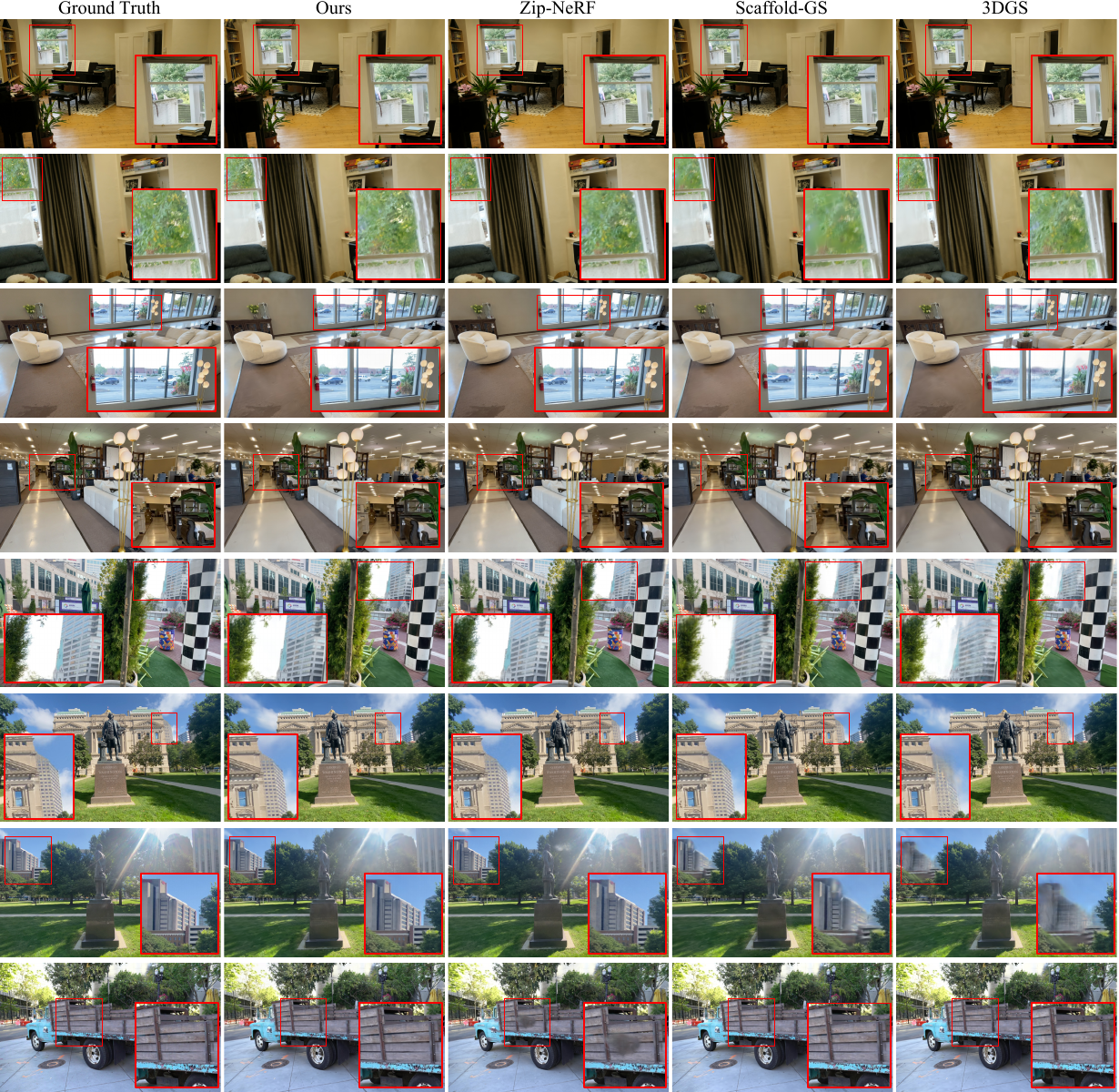}\vspace{-2mm}
   \caption{
   \textbf{Extra visual comparisons on novel view synthesis 1.} We present visual comparisons between methods on more test views. The figure includes ROOM from the Mip-NeRF360 dataset, TRUCK from Tanks\&Temples, and scenes 21, 26, and 69 from the DL3DV benchmark. Key differences in quality are highlighted by insets.}\vspace{-2mm}
   \label{fig:supp_main1}
\end{figure*}
\begin{figure*}[tp]
  \centering
  \includegraphics[width=\linewidth]{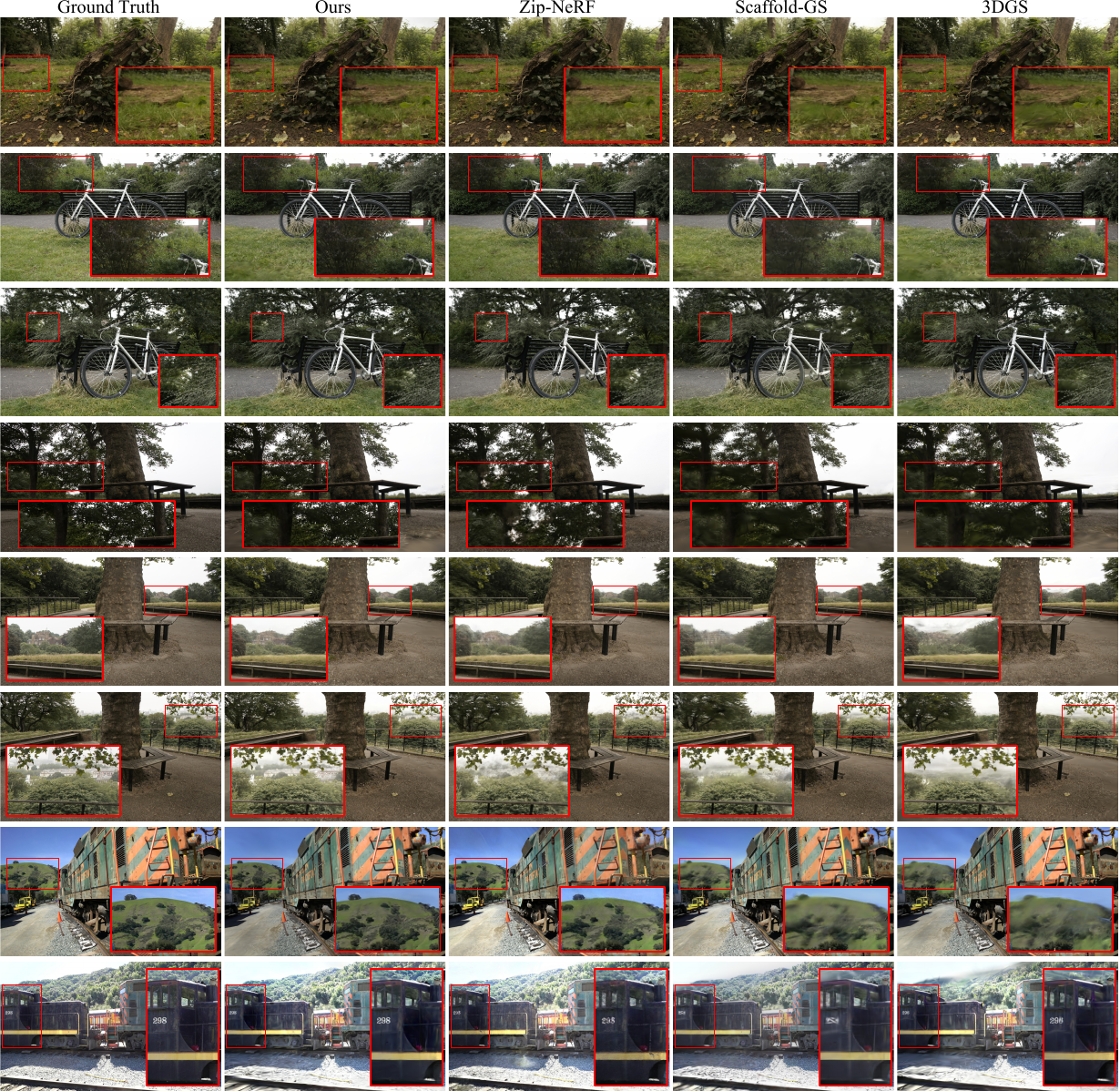}\vspace{-2mm}
   \caption{
   \textbf{Extra visual comparisons on novel view synthesis 2.} We present visual comparisons between methods on more test views. The figure includes STUMP, BICYCLE, and TREEHILL from the Mip-NeRF360 dataset; and TRAIN from Tanks\&Temples. Key differences in quality are highlighted by insets.}\vspace{-2mm}
   \label{fig:supp_main2}
\end{figure*}

\section{Additional Results}
\label{supp:additional}

We provide quantitative comparisons per scene in Tables~\ref{tab:360scenes}--\ref{tab:dl3dv_outdoor},
and qualitative results in Figs.~\ref{fig:supp_main1} and \ref{fig:supp_main2} to supplement the main paper. These include detailed figures and tables showcasing the performance of our method across various scenes.

\subsection{Distribution of $w$}
\label{supp:distribution_w}
We provide a histogram of the parameter \(w\) and the mean distance for each \(w\) on the TRAIN scene from Tanks~\&~Temples \cite{tanks}. indicates that \(w\) correctly converges to small values (\ie, $w\sim0$) for distant scenes, and \textit{vice versa}.

\subsection{Comparison with Mip-Splatting and Multi-Scale 3D Gaussian Splatting}
\label{supp:compare_extra}
We provide a comparison with 3DGS-based methods~\cite{Yu2024MipSplatting, yan2024multiscale3dgaussiansplatting} that focus on anti-aliasing and multi-scale representations. Even with these methods, our method still maintains state-of-the-art performance across most metrics.

\end{document}